\documentclass[10pt]{article}
\usepackage{typearea}\typearea{14}
\usepackage{epsfig,amsmath,amsfonts,amssymb,upgreek,dsfont}
\newtheorem{theorem}{Theorem}[section]

\makeatletter
\@addtoreset{equation}{section}
\makeatother

\makeatletter
\long\def\@makecaption#1#2{{\small
\advance\leftskip1cm
\advance\rightskip1cm
\vskip\abovecaptionskip
\sbox\@tempboxa{#1: #2}%
\ifdim \wd\@tempboxa >\hsize
 #1: #2\par
\else
\global \@minipagefalse
\hb@xt@\hsize{\hfil\box\@tempboxa\hfil}%
\fi
\vskip\belowcaptionskip}}
\makeatother
\def\eq#1\en{\begin{equation}#1\end{equation}}  
\def\eqa#1\ena{\begin{align}#1\end{align}}
\def\eqg#1\eng{\begin{gather}#1\end{gather}}
\newcommand{\lb}[1]{\label{e:#1}}
\newcommand{\rlb}[1]{\eqref{e:#1}} 
\newcommand{\nl}{\notag\\}

\newcommand{\sbkt}[1]{\langle#1\rangle}

\newcommand{\sumtwo}[2]%
{\mathop{\sum_{#1}}_{#2}}
\newcommand{\sumthree}[3]%
{\mathop{\mathop{\sum_{#1}}_{#2}}_{#3}}
\newcommand{\sumfour}[4]%
{\mathop{\mathop{\mathop{\sum_{#1}}_{#2}}_{#3}}_{#4}} 
\newcommand{\prodtwo}[2]%
{\mathop{\prod_{#1}}_{#2}}
\newcommand{\mintwo}[2]%
{\mathop{\min_{#1}}_{#2}}
\newcommand{\maxtwo}[2]%
{\mathop{\max_{#1}}_{#2}}
\newcommand{\maxthree}[3]%
{\mathop{\mathop{\max_{#1}}_{#2}}_{#3}}
\newcommand{\limtwo}[2]%
{\mathop{\lim_{#1}}_{#2}}
\newcommand{\suptwo}[2]%
{\mathop{\sup_{#1}}_{#2}}
\newcommand{\supthree}[3]%
{\mathop{\mathop{\sup_{#1}}_{#2}}_{#3}}
\newcommand{\supfour}[4]%
{\mathop{\mathop{\mathop{\sup_{#1}}_{#2}}_{#3}}_{#4}} 
\newcommand{\inftwo}[2]%
{\mathop{\inf_{#1}}_{#2}}
\newcommand{\infthree}[3]%
{\mathop{\mathop{\inf_{#1}}_{#2}}_{#3}}
\newcommand{\inffour}[4]%
{\mathop{\mathop{\mathop{\inf_{#1}}_{#2}}_{#3}}_{#4}} 

\newcommand\calM{{\cal M}}

\newcommand\calS{{\cal S}}

\newcommand{\mrx}{\mathrm{x}}
\newcommand{\mry}{\mathrm{y}}
\newcommand{\mrz}{\mathrm{z}}

\newcommand{\sfA}{\mathsf{A}}
\newcommand{\sfB}{\mathsf{B}}

\newcommand{\sfD}{\mathsf{D}}

\newcommand{\sfI}{\mathsf{I}}

\newcommand{\sfO}{\mathsf{O}}

\newcommand{\sfT}{\mathsf{T}}

\newcommand{\bbZ}{\mathbb{Z}}

\newcommand{\up}{\uparrow}
\newcommand{\dn}{\downarrow}

\newcommand{\Di}{\mathit{\Delta}}
\newcommand{\qedm}{\rule{1.5mm}{3mm}}

\newcommand{\hS}{\hat{S}}
\newcommand{\hK}{\hat{K}}
\newcommand{\hM}{\hat{M}}
\newcommand{\hA}{\hat{A}}
\newcommand{\hB}{\hat{B}}
\newcommand{\hH}{\hat{H}}
\newcommand{\hO}{\hat{O}}
\newcommand{\hP}{\hat{P}}
\newcommand{\hQ}{\hat{Q}}
\newcommand{\hU}{\hat{U}}

\newcommand{\hh}{\hat{h}}

\newcommand{\hSx}{\hat{S}^{(\mrx)}}
\newcommand{\hSy}{\hat{S}^{(\mry)}}
\newcommand{\hSz}{\hat{S}^{(\mrz)}}

\newcommand{\hUy}{\hat{U}^{(\mry)}}
\newcommand{\hUz}{\hat{U}^{(\mrz)}}

\newcommand{\kone}{\ket{1}}
\newcommand{\kzero}{\ket{0}}
\newcommand{\kmone}{\ket{\!-\!1}}

\newcommand{\bra}[1]{\langle#1|}
\newcommand{\ket}[1]{|#1\rangle}
\newcommand{\Tr}{\operatorname{Tr}}
\newcommand{\pmat}[1]{\begin{pmatrix}#1\end{pmatrix}}

\newcommand{\ZZ}{\bbZ_2\times\bbZ_2}

\newcommand{\pmu}[1]{\ket{\psi^{(\mu)}_{#1}}}
\newcommand{\uket}{\ket{\!\!\up}}
\newcommand{\dket}{\ket{\!\!\dn}}
\newcommand{\ubra}{\bra{\up\!\!}}
\newcommand{\dbra}{\bra{\dn\!\!}}

\newcommand{\hT}{\hat{\Theta}}
\newcommand{\Sy}{\hat{\calS}}

\newcommand{\ra}{\mathrm{a}}
\newcommand{\rb}{\mathrm{b}}
\newcommand{\rc}{\mathrm{c}}
\newcommand{\rd}{\mathrm{d}}
\newcommand{\rx}{\mathrm{x}}
\newcommand{\ry}{\mathrm{y}}
\newcommand{\rz}{\mathrm{z}}

\newcommand{\lamax}{\lambda_\mathrm{max}}

\begin{document}

\noindent
{\Large\bf The Asymmetric Valence-Bond-Solid States in Quantum Spin Chains}

\vspace{0.8mm}

\noindent{\large\bf
The Difference Between Odd and Even Spins}

\renewcommand{\thefootnote}{\fnsymbol{footnote}}
\medskip\noindent
Daisuke Maekawa and Hal Tasaki\footnote{%
Department of Physics, Gakushuin University, Mejiro, Toshima-ku, 
Tokyo 171-8588, Japan.
}
\renewcommand{\thefootnote}{\arabic{footnote}}
\setcounter{footnote}{0}

\begin{quotation}
\small\noindent
The qualitative difference in low-energy properties of spin $S$ quantum antiferromagnetic chains with integer $S$ and half-odd-integer $S$ discovered by Haldane can be intuitively understood in terms of the valence-bond picture proposed by Affleck, Kennedy, Lieb, and Tasaki.
Here we develop a similarly intuitive diagrammatic explanation of the qualitative difference between chains with odd $S$ and even $S$, which is at the heart of the theory of symmetry-protected topological (SPT) phases.

More precisely, we define one-parameter families of states, which we call the asymmetric valence-bond solid (VBS) states, that continuously interpolate between the Affleck-Kennedy-Lieb-Tasaki (AKLT) state and the trivial zero state in quantum spin chains with $S=1$ and 2.
The asymmetric VBS state is obtained by systematically modifying the AKLT state.
It always has exponentially decaying truncated correlation functions and is a unique gapped ground state of a short-ranged Hamiltonian.
We also observe that the asymmetric VBS state possesses the time-reversal, the $\ZZ$, and the bond-centered inversion symmetries for $S=2$, but not for $S=1$.
This is consistent with the known fact that the AKLT model belongs to the trivial SPT phase if $S=2$ and to a nontrivial SPT phase if $S=1$.
Although such interpolating families of disordered states were already known, our construction is unified and is based on a simple physical picture.
It also extends to spin chains with general integer $S$ and provides us with an intuitive explanation of the essential difference between models with odd and even spins. 

There is a 24 minutes video in which the essence of the present work is discussed:
\newline
\url{https://youtu.be/URsf9e_PLlc}
\end{quotation}

\tableofcontents

\section{Introduction}
\label{s:intro}
In the early 1980s, Haldane made the now-famous discovery that there is a qualitative difference in low energy behaviors of the spin $S$ antiferromagnetic Heisenberg chain with half-odd integer $S$ and integer $S$ \cite{Haldane1981,Haldane1983a,Haldane1983b}.
See, e.g., Part~II of \cite{TasakiBook} for a review.
It was shown that the ground state of the model is critical for half-odd integer $S$ while that for integer $S$ is disordered.
Here we say that a ground state is critical if it is a unique ground state accompanied by gapless excitations, and the corresponding truncated correlation functions decay by power laws.
We say that a ground state is disordered if it is a unique gapped ground state, i.e., a unique ground state accompanied by a nonzero energy gap.\footnote{
We note that our use of the term ``disordered'' may sometimes not be consistent with one's intuition.
Consider the Hamiltonian $\hH=-\sum_j\hSz_j$, which describes independent spins in a uniform magnetic field.
The ground state $\ket{\Phi_\text{up}}=\bigotimes_j\ket{S}_j$ is obviously unique, accompanied by a gap, and has vanishing truncated correlations.
We thus call $\ket{\Phi_\text{up}}$ a disordered ground state although all the spins are pointing in the same direction.
}
In such a state, it is known that truncated correlation functions always decay exponentially \cite{HastingsKoma,NS1}.

Although the qualitative difference between models with half-odd-integer $S$ and integer $S$ appears mysterious, there are some convincing theoretical arguments that support it.
In particular, it is known that the difference can be intuitively understood from diagrams of ``valence-bonds" as in Figures~\ref{f:VBS1} and \ref{f:VBS2}.
This valence-bond picture is a byproduct of exactly solvable spin chains, now called the Affleck-Kennedy-Lieb-Tasaki (AKLT) model, which resembles the Heisenberg antiferromagnet and is proved to have a unique gapped ground state \cite{AKLT1,AKLT2,FannesNachtergaeleWerner1992}.

Subsequent studies of quantum spin chains triggered by Haldane's discovery had led to the notion of hidden antiferromagnetic order \cite{denNijsRommelse} and accompanying hidden symmetry breaking \cite{KennedyTasaki1992A,KennedyTasaki1992B} in $S=1$ antiferromagnetic chains.
In 1992, Oshikawa realized that there is a further qualitative difference between models with odd $S$ and even $S$ \cite{Oshikawa92}.
It was found that the unique gapped ground state of spin $S$ antiferromagnetic chain with odd $S$ has a hidden order while that with even $S$ has no hidden order. 
This peculiar phenomenon was not fully understood until 2009 when Gu and Wen \cite{GuWen2009} and then Pollmann, Turner, Berg, and Oshikawa \cite{PollmannTurnerBergOshikawa2010,PollmannTurnerBergOshikawa2012} established the notion of symmetry-protected topological (SPT) phases.
In particular, the qualitative difference between models with odd and even $S$ was discussed and essentially resolved by Pollmann, Turner, Berg, and Oshikawa in \cite{PollmannTurnerBergOshikawa2012}.\footnote{%
A preprint version of this paper appeared in arXiv in 2009, slightly after \cite{GuWen2009}.
}

To see this, let us briefly discuss the main idea of SPT phases in the context of quantum spin chains with fixed spin $S$. 
Let $\calM$ be the (big) space of all Hamiltonians with short-ranged and (uniformly) bounded interactions that have a unique gapped ground state. 
(To be precise, we are treating spin systems on the infinite chain.
See section~\ref{s:BG}.)
We say that two Hamiltonians $\hH_0$ and $\hH_1$ in $\calM$ are continuously connected if there exists a one-parameter family of Hamiltonians $\hH_s\in\calM$ with $s\in[0,1]$ that depends continuously on $s$.  
It is believed that any $\hH_0$ and $\hH_1$ in $\calM$ are continuously connected, and hence all Hamiltonians in $\calM$ belong to a single ``phase'' \cite{ChenGuWEn2011}. This fact was proved for matrix product states by Ogata \cite{Ogata1,Ogata2,Ogata3}.

The situation changes drastically if one considers Hamiltonians that possess certain global symmetry $\Sigma$.
Let $\calM_\Sigma$ be the subspace of $\calM$ that consists of Hamiltonians with symmetry $\Sigma$. 
We say that two Hamiltonians $\hH_0$ and $\hH_1$ in $\calM_\Sigma$ are continuously connected within $\calM_\Sigma$ if there exists a one-parameter family of Hamiltonians $\hH_s\in\calM_\Sigma$ with $s\in[0,1]$ that depends continuously on $s$.  
Depending on the choice of symmetry $\Sigma$, it may happen that $\calM_\Sigma$ is divided into several distinct connected components.  
These components are called topological phases protected by the symmetry $\Sigma$, or, simply, SPT phases.
It is known that, when $\Sigma$ is the time-reversal symmetry, the $\ZZ$ symmetry, or the bond-centered inversion symmetry, the corresponding space $\calM_\Sigma$ for a spin chain with integer $S$ consists of at least two SPT phases. 
One of the phases contains a trivial product state and hence is called the trivial phase.  
We should note that a mathematically rigorous and general theory of SPT phases in quantum spin chains was recently established by Ogata \cite{OgataZ2,Ogatainv,OgataCDM,HalSPT}.

Given the picture of SPT phases, the difference between odd $S$ and even $S$ is that the AKLT model belongs to a nontrivial SPT phase when $S$ is odd and to the trivial phase when $S$ is even.  
In other words, the AKLT Hamiltonian is continuously connected to a trivial Hamiltonian within $\calM_\Sigma$ if $S$ is even but not connected when $S$ is odd.

The main goal of the present paper is to provide an intuitive picture that explains the difference between spin chains with odd $S$ and even $S$. 
We shall see that the difference can be understood diagrammatically in terms of  ``asymmetric valence-bonds'' as in Figure~\ref{f:mVBS3}.
To be more precise, we construct simple one-parameter families of disordered states\footnote{%
We say that a state is disordered if all truncated correlation functions decay exponentially.} for $S=1$ and 2 spin chains that connect a trivial product state and the AKLT state, i.e., the ground state of the AKLT model. 
We call these states the asymmetric valence-bond-solid (VBS) states.
The asymmetric VBS states are obtained from the AKLT state by systematically replacing the valence-bond, which is the basic building block of the AKLT state, with a state we call the asymmetric valence-bond.
The asymmetric valence-bond is similar to the valence-bond, i.e., the spin-singlet, and is characterized by an asymmetry parameter $\mu\in[0,1]$.
As far as we know, this particular one-parameter modification of the AKLT state has not been discussed in the literature.
It is interesting to find other examples where similar modification leads to meaningful states and models.
See also section~\ref{s:MO} and Appendix~\ref{s:Witten}.

We shall see that the $S=2$ asymmetric VBS state possesses the time-reversal, the $\ZZ$, and the bond-centered inversion symmetries, while the $S=1$ state does not have any of these symmetries.
Note that this is consistent with the above picture of the SPT phases.
We also show that there exist continuous families of Hamiltonians that have the asymmetric VBS states as their unique gapped ground states.
We note that similar families of disordered states (and the corresponding Hamiltonians) that connect a trivial state and the AKLT state were already constructed by Bachman and Nachtergale for $S=1$ \cite{BachmannNachtergaele2012,BachmannNachtergaele2014} and by Pollmann, Turner, Berg, and Oshikawa for $S=2$ \cite{PollmannTurnerBergOshikawa2012}.
The advantage of our approach is that the states we propose are simpler and can be understood intuitively.
Moreover, our construction is unified and extends naturally to models with general integer spin $S$, thus clarifying the essential difference between even and odd $S$.

\section{Background}
\label{s:BG}
Let us make the preceding discussion concrete and set our goal.

We start by fixing general notations.
See, e.g., Chapter~2 of \cite{TasakiBook} for basics about quantum spin systems.
The spin operators of a single spin are denoted as $(\hSx,\hSy,\hSz)$.
They are $(2S+1)\times(2S+1)$ matrices and satisfy $(\hSx)^2+(\hSy)^2+(\hSz)^2=S(S+1)\,\hat{1}$, where $S=\frac{1}{2},1,\frac{2}{3},\ldots$ is the spin quantum number.
We use the standard basis state $\ket{m}$ such that $\hSz\ket{m}=m\ket{m}$, where $m=-S,-S+1,\ldots,S$.
When $S=\frac{1}{2}$, we write $\uket$ and $\dket$ instead of $\ket{\frac{1}{2}}$ and $\ket{-\frac{1}{2}}$, respectively.
For a spin system on a finite lattice, we denote by $(\hSx_p,\hSy_p,\hSz_p)$ and $\ket{m}_p$ the spin operators and the basis states corresponding to a site $p$.

Consider a finite one-dimensional lattice whose sites are denoted as $j=1,\ldots,L$.  
Unless otherwise mentioned, we take the periodic boundary condition and identify the site $L+1$ with 1.
We assume there is a quantum spin with the spin quantum number $S$ on each site $j$.
The standard basis state of the spin chain is 
\eq
\bigotimes_{j=1}^L\ket{m_j}_j,
\lb{basism}
\en
where $m_j=-S,\-S+1,\ldots,S$.

In this paper, we concentrate on spin chains with a short-ranged translation-invariant Hamiltonian
\eq
\hH=\sum_{j=1}^L\hh_j,
\lb{Hgen}
\en
where $\hh_j$ is the translation of a Hermitian operator $\hh_1$ that acts nontrivially only on a finite number of sites including $j=1$.
We say that the spin chain with Hamiltonian $\hH$ has a unique gapped ground state if there exists an increasing sequence $L_1,L_2,\ldots$ of positive integers\footnote{An example is the sequence $2,4,\ldots$ of even numbers.} such that (i)~for any $i=1,2,\ldots$, the Hamiltonian on the chain with $L_i$ sites has a unique ground state accompanied by an energy gap that is not less than $\Di E$, where $\Di E>0$ is independent of $i$, and (ii)~the $i\up\infty$ limit of the finite chain ground states defines a state $\omega$ on the infinite chain.\footnote{%
To be precise, for each $i$, we define the ground state $\ket{\Phi^{(i)}_{\rm GS}}$ on the finite chain $\{-(L_i/2)+1,\ldots,L_i/2\}$ by shifting the original ground state on $\{1,\ldots,L_i\}$.
Then the state $\omega$ is defined by $\omega(\hA)=\lim_{i\up\infty}\bra{\Phi^{(i)}_{\rm GS}}\hA\ket{\Phi^{(i)}_{\rm GS}}$ for any local operator $\hA$.
See, e.g., \cite{TasakiBook,TasakiLSM} for more details.
}
Then it can be shown that the state $\omega$ is a locally-unique gapped ground state of the infinite chain with (the formal infinite volume limit of) the Hamiltonian $\hH$.
Moreover, when the Hamiltonian $\hH$ has relevant symmetry, one can associate the ground state $\omega$ with a unique topological index, i.e., the Ogata index \cite{OgataZ2,OgataCDM}, that characterizes the SPT phases.
See \cite{TasakiLSM}.

Haldane \cite{Haldane1981,Haldane1983a,Haldane1983b} studied standard one-dimensional spin systems including the spin $S$ antiferromagnet Heisenberg chain with the Hamiltonian
\eq
\hH_\mathrm{Heis}=\sum_{j=1}^L\hat{\boldsymbol{S}}_j\cdot\hat{\boldsymbol{S}}_{j+1},
\en
where $L$ is even, and argued that the model has a unique gapped ground state when $S$ is an integer.
This conclusion is not yet rigorously proved for standard models but is known to be valid for the Affleck-Kennedy-Lieb-Tasaki (AKLT) model with the Hamiltonian
\eq
\hH_\mathrm{AKLT}=\sum_{j=1}^L\hP^{S_\mathrm{tot}>S}_{j,j+1},
\lb{HAKLT}
\en
where $\hP^{S_\mathrm{tot}>S}_{j,j+1}$ denotes the projection operator onto the sector in which the total spin at two sites $j$ and $j+1$ is larger than $S$ \cite{AKLT1,AKLT2}.\footnote{Recall that the total spin $S_\mathrm{tot}$ of two spins with quantum number $S$ takes the values $0,1,\ldots,2S$.}
In the original work \cite{AKLT1,AKLT2}, it was proved that the $S=1$ chain with the Hamiltonian \rlb{HAKLT} has a unique gapped ground state.
The same conclusion for the chain with general integer $S$ was later proved by Fannes, Nachtergaele, and Werner as an application of the general theory of matrix product states.
See section~7.3 of \cite{FannesNachtergaeleWerner1992}.
The ground states of the AKLT model, which are called the AKLT states or the valence-bond solid (VBS) states, are compactly expressed in terms of valence-bonds, i.e., spin-singlets\footnote{
We here define valence-bond as an unnormalized state $\uket_p\dket_q-\dket_p\uket_q$.
} formed by two $S=\frac{1}{2}$ spins, and projection operators as
\eq
\ket{\Phi_\mathrm{AKLT}}=\Bigl(\bigotimes_{j=1}^L\Sy_j\Bigr)\Bigl(\bigotimes_{j=1}^L\bigl\{\uket_{\!(j,\rb)}\dket_{\!(j+1,\ra)}-\dket_{\!(j,\rb)}\uket_{\!(j+1,\ra)}\bigr\}\Bigr),
\lb{AKLT1}
\en
for $S=1$, and 
\eq
\ket{\Phi_\mathrm{AKLT}}=\Bigl(\bigotimes_{j=1}^L\Sy_j\Bigr)\Bigl(\bigotimes_{j=1}^L
\bigl\{\uket_{\!(j,\rb)}\dket_{\!(j+1,\ra)}-\dket_{\!(j,\rb)}\uket_{\!(j+1,\ra)}\bigr\}\otimes
\bigl\{\uket_{\!(j,\rd)}\dket_{\!(j+1,\rc)}-\dket_{\!(j,\rd)}\uket_{\!(j+1,\rc)}\bigr\}
\Bigr),
\lb{AKLT2}
\en
for $S=2$.
See sections~\ref{s:S=1} and \ref{s:S=2}, respectively, for the notation.

\begin{figure}
\centerline{\epsfig{file=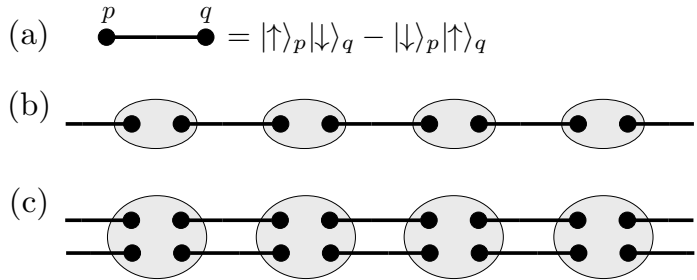,width=9truecm}}
\caption[dummy]{
In all the figures in the present paper, a black dot represents a spin with $S=\frac{1}{2}$.
(a)~Two black dots connected by a solid line stands for the valence-bond, i.e., the spin-singlet formed by two spins.
(b)~The diagrammatic representation of the $S=1$ AKLT state \rlb{AKLT1}.
Two black dots, i.e., two spin $\frac{1}{2}$'s, surrounded by an oval denote the symmetrized state, which is a state of $S=1$.
See \rlb{S} for the precise definition.
(c)~The diagrammatic representation of the $S=2$ AKLT state \rlb{AKLT2}.
In this case, the symmetrization of four spin $\frac{1}{2}$'s gives an $S=2$ spin.
}
\label{f:VBS1}
\end{figure}

\begin{figure}
\centerline{\epsfig{file=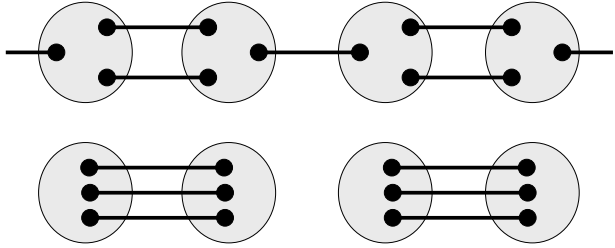,width=8truecm}}
\caption[dummy]{
States of the $S=3/2$ chain that can be constructed by using valence-bonds of unit length.
Since there must be three valence-bonds attached to each site, one inevitably gets non-translationally-invariant states.
}
\label{f:VBS2}
\end{figure}

The AKLT states have simple diagrammatic representations as in Figure~\ref{f:VBS1}.
Note that there are two valence-bonds attached to each site in (a), which means that each site carries two spin $\frac{1}{2}$'s forming a spin with $S=1$.
Similarly, there are four valence-bonds attached to each site in (b), which correspond to having  $S=2$ spin at each site.
One can draw similar diagrams for the AKLT state with any integer $S$, in which each pair of neighboring sites carries $S$ valence-bonds.
When $S$ is a half-odd integer, on the other hand, each site should consist of an odd number (which is $2S$) of spin $\frac{1}{2}$'s.
Forming a state in terms of valence-bonds with limited length is impossible without breaking translation invariance. 
See Figure~\ref{f:VBS2}.
One thus sees a clear qualitative difference between spin chains with integer $S$ and half-odd integer $S$, i.e., one can form a translation-invariant state by using short-ranged valence-bonds only when $S$ is an integer.
This is the valence-bond picture that explains Haldane's discovery.

Next, consider the trivial Hamiltonian
\eq
\hH_\mathrm{trivial}=\sum_{j=1}^L(\hSz_j)^2.
\lb{Htrivial}
\en
When $S$ is an integer, the Hamiltonian has a unique gapped ground state
\eq
\ket{\Phi_\mathrm{zero}}=\bigotimes_{j=1}^L\ket{0}_j,
\lb{zero}
\en
which we call the zero state.
Note that the truncated correlation function of the zero state is vanishing.

As we discussed in section~\ref{s:BG}, we shall focus on the question of whether the AKLT model \rlb{HAKLT} is continuously connected to the trivial model \rlb{Htrivial} within the whole space $\calM$ or within the restricted space $\calM_\Sigma$ for the three types of symmetry $\Sigma$.
Our strategy is to explicitly construct disordered states, which we call the asymmetric VBS states, that continuously interpolate between the AKLT state and the zero state.

Finally, we shall be more specific about the types of symmetry.
For SPT phases in quantum spin chains, the relevant symmetries are those with respect to (i)~the time-reversal transformation, (ii)~the $\ZZ$ transformation, and (iii)~the bond-centered inversion transformation.
Let us briefly discuss the three symmetries.
See, e.g., \cite{TasakiBook} for details.

\medskip\par
(i)~The time-reversal transformation for a single spin with $S=m$ is described by the antiunitary operator $\hT$ that acts on the  basis states as
\eq
\hT\ket{m}=(-1)^{S-m}\ket{-m}.
\lb{Tm}
\en
When $S=\frac{1}{2}$, we have
\eq
\hT\uket=\dket,\quad\hT\dket=-\uket.
\lb{T}
\en
The time-reversal transformation of the spin chain, which we denote by the same symbol $\hT$, is defined as the antiunitary operator that acts on basis state on each site as \rlb{Tm} or \rlb{T}.

\medskip\par
(ii)~The $\ZZ$ transformation is generated by the global $\pi$-rotations of the spins about two mutually orthogonal axes, e.g., the y-axis and the z-axis.
They are described by the unitary operators $\hUy=\exp[-i\pi\sum_{j=1}^L\hSy_j]$ and $\hUz=\exp[-i\pi\sum_{j=1}^L\hSz_j]$.

\medskip\par
(iii)~The bond-centered inversion transformation is the unitary transformation induced by the inversion transformation $j\to L+1-j$.
(Note that this inversion is centered at the bond $(L,1)$.)
It acts on the basis state \rlb{basism} as 
\eq
\hU_{\rm inv}\bigotimes_{j=1}^L\ket{m_j}_j=\bigotimes_{j=1}^L\ket{m_{L+1-j}}_j.
\lb{basismU}
\en

\section{Asymmetric valence-bond}
\label{s:mu}
Let us define the asymmetric valence-bond\footnote{
We note this is a slight abuse of terminology since a valence-bond should not be asymmetric.
}, and discuss its basic transformation properties.
The asymmetric valence-bond is the building block of the asymmetric VBS state.

\begin{figure}
\centerline{\epsfig{file=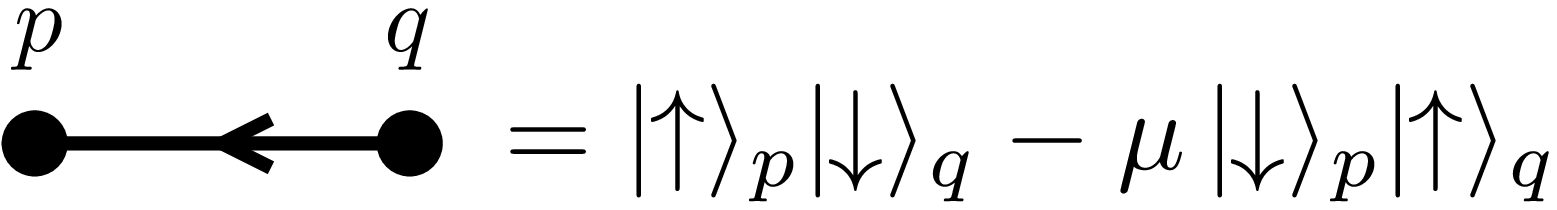,width=5truecm}}
\caption[dummy]{
The asymmetric valence-bond $\pmu{p,q}$ that consists of two $S=\frac{1}{2}$'s at sites $p$ and $q$.
Unlike the spin-singlet, the state $\pmu{p,q}$ is not invariant if one switches $p$ and $q$.
We thus represent the asymmetric valence-bond diagrammatically by a line with an arrow.
}
\label{f:mVB}
\end{figure}

Consider a system of two sites, which we call $p$ and $q$, and associate each site with a spin with $S=\frac{1}{2}$.
For $\mu$ such that $0\le\mu\le1$, we define the asymmetric valence-bond as
\eq
\pmu{p,q}=\uket_p\dket_q-\mu\,\dket_p\uket_q.
\lb{mu}
\en
Note that the state $\pmu{p,q}$ with $\mu=1$ is nothing but the spin-singlet or the standard valence-bond.
Although the spin-singlet is invariant under the inversion $p\leftrightarrow q$ in the sense that $\ket{\psi^{(1)}_{p,q}}=-\ket{\psi^{(1)}_{q,p}}$, the state $\pmu{p,q}$ with $\mu<1$ does not have such (bond-centered) inversion symmetry.
We thus represent the asymmetric valence-bond diagrammatically by an oriented segment as in Figure~\ref{f:mVB}.

From \rlb{mu} and \rlb{T} we find that the asymmetric valence-bond transforms under the time-reversal as
\eq
\hT\pmu{p,q}=-\pmu{q,p},
\lb{Tp}
\en
where one should note that $p$ and $q$ are switched on the right-hand side.
Recalling the above remark about the inversion invariance, one finds that the state is not time-reversally invariant unless $\mu=1$.
Likewise, we find that the state transforms under the $\ZZ$ transformation as
\eq
\hUy\pmu{p,q}=-\pmu{q,p},\quad\hUz\pmu{p,q}=\pmu{p,q}.
\lb{Up}
\en
Again the asymmetric valence-bond $\pmu{p,q}$ is not $\ZZ$ invariant unless $\mu=1$.

\section{The asymmetric VBS state for the $S=1$ chain}
\label{s:S=1}
We consider an $S=1$ system on the chain with $L$ sites (where $L$ is an arbitrary positive integer) with the periodic boundary condition and define the asymmetric VBS state.
We shall show that the asymmetric VBS state continuously connects the AKLT state \rlb{AKLT1} and the zero state \rlb{zero}, but lacks any of the three relevant symmetries.
We note that our state is similar to those constructed and discussed by Bachmann and Nachtergaele in \cite{BachmannNachtergaele2012} and section~3 of \cite{BachmannNachtergaele2014} in order to connect the AKLT state and the zero state.
We also find that the asymmetric VBS state can be regarded as a particular case of the general matrix product states constructed in \cite{KLZ}.
It is similar to but different from the $q$-deformed AKLT state, i.e., Example 7 of \cite{FannesNachtergaeleWerner1992}.

\subsection{The state and the (lack of) symmetry}
We follow the original construction of the AKLT state \cite{AKLT1,AKLT2} and represent a spin with $S=1$ in terms of two spins with $S=\frac{1}{2}$.
Consider two $S=\frac{1}{2}$ spins labeled by a and b.
We define the projection operator onto the triplet as
\eq
\Sy=\kone\,{}_\rb\!\ubra_\ra\!\ubra+\kzero\frac{{}_\rb\!\dbra_\ra\!\ubra+{}_\rb\!\ubra_\ra\!\dbra}{\sqrt{2}}+\kmone\,{}_\rb\!\dbra_\ra\!\dbra,
\lb{S}
\en
where $\ket{m}$ with $m=0,\pm1$ denote the standard basis states of a spin with $S=1$.
Note that $\Sy$ can be interpreted as the symmetrization operator \cite{AKLT1,AKLT2,TasakiBook}.

\begin{figure}
\centerline{\epsfig{file=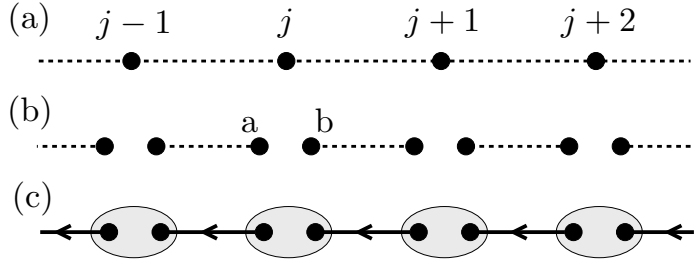,width=9truecm}}
\caption[dummy]{
The construction of the $S=1$ asymmetric VBS state.
(a)~The original chain with $L$ sites.
(b)~The duplicated chain.  We doubled a single site $j$ into two sites $(j,\ra)$ and $(j,\rb)$.
(c)~The $S=1$ asymmetric VBS state. 
A gray oval represents the symmetrization of two spin $\frac{1}{2}$'s, which gives an $S=1$.
}
\label{f:mVBS1}
\end{figure}

We consider the duplicated chain obtained by doubling each site in the original chain, and label doubled sites as $(j,\ra)$ and $(j,\rb)$ with $j=1,\ldots,L$.
We associate each site in the duplicated chain with an $S=\frac{1}{2}$ spin.
Then we define the $S=1$ asymmetric VBS state characterized by the parameter $\mu\in[0,1]$ as
\eq
\ket{\Psi_\mu}=\Bigl(\bigotimes_{j=1}^L\Sy_j\Bigr)\Bigl(\bigotimes_{j=1}^L\pmu{(j,\rb),(j+1,\ra)}\Bigr),
\lb{mu1}
\en
where $\pmu{(j,\rb),(j+1,\ra)}$ is the asymmetric valence-bond \rlb{mu} and $\Sy_j$ is the operator \rlb{S} that acts on the two spins on sites $(j,\ra)$ and $(j,\rb)$.
See Figure~\ref{f:mVBS1}.

Clearly the asymmetric VBS state \rlb{mu1} coincides with the AKLT state \rlb{AKLT1} when $\mu=1$.
For $\mu=0$, we see from \rlb{mu} and \rlb{S} that
\eq
\ket{\Psi_0}=\Bigl(\bigotimes_{j=1}^L\Sy_j\Bigr)\Bigl(\bigotimes_{j=1}^L\uket_{\!(j,\rb)}\dket_{\!(j+1,\ra)}\Bigr)
=2^{-L/2}\bigotimes_{j=1}^L\kzero_j=2^{-L/2}\,\ket{\Phi_\mathrm{zero}}.
\lb{P0}
\en
We thus see that the asymmetric VBS state $\ket{\Psi_\mu}$ continuously interpolates between the AKLT state  \rlb{AKLT1} and the zero state \rlb{zero}.

From \rlb{Tp} and \rlb{Up}, one easily sees that the asymmetric VBS state $\ket{\Psi_\mu}$ with $0<\mu<1$ is not invariant under the time-reversal, the $\ZZ$, or the bond-centered inversion transformation.
Recall that the AKLT state $\ket{\Psi_1}$ is invariant under these three transformations.
Although the asymmetric valence-bond $\pmu{(j,\rb),(j+1,\ra)}$ with $\mu=0$ lacks the symmetry, the zero state \rlb{P0} as a whole is invariant under the three transformations.

\subsection{The matrix product representation and correlation functions}
Recalling that $\pmu{(j,\rb),(j+1,\ra)}$ is a sum of two terms as in \rlb{mu}, one can expand the asymmetric VBS state \rlb{mu1} into a sum of $2^L$ terms.
The expansion is neatly expressed in the matrix product form
\eq
\ket{\Psi_\mu}=\sum_{m_1,\ldots,m_L=-S}^S\Tr[\sfA^{(m_1)}\cdots\sfA^{(m_L)}]\,
\ket{m_1}_1\otimes\cdots\otimes\ket{m_L}_L,
\lb{mps}
\en
where we set $S=1$ in the present section.
It is found that the matrices in \rlb{mps} may be chosen as
\eq
\sfA^{(1)}=\pmat{0&0\\-\sqrt{\mu}&0},\quad
\sfA^{(0)}=\frac{1}{\sqrt{2}}\pmat{1&0\\0&-\mu},\quad
\sfA^{(-1)}=\pmat{0&\sqrt{\mu}\\0&0}.
\lb{A1}
\en
See, e.g., section~7.2.2 of \cite{TasakiBook} for details.

\begin{figure}
\centerline{\epsfig{file=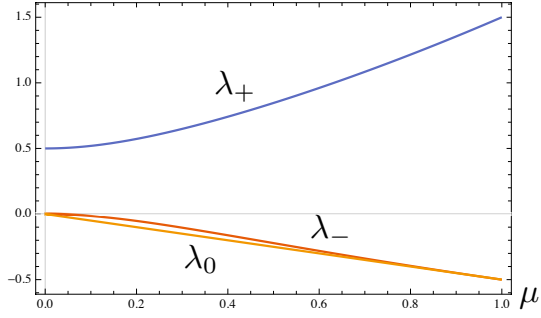,width=7truecm}}
\caption[dummy]{
The eigenvalues \rlb{lambda1}.
They are ordered as $\lambda_+>0>\lambda_->\lambda_0$ for $\mu\in(0,1)$.
}
\label{f:mevs1}
\end{figure}

From the matrix product representation \rlb{mps}, one finds that the normalization of the asymmetric VBS state is expressed as 
\eq
\sbkt{\Psi_\mu|\Psi_\mu}=\Tr[\sfT^L],
\lb{norm}
\en
where the transfer matrix $\sfT$ is the $4\times4$ matrix defined as
\eq
(\sfT)_{(\alpha,\beta),(\alpha',\beta')}:=
\sum_{m=-S}^S(\sfA^{(m)})_{\alpha,\alpha'}(\sfA^{(m)})_{\beta,\beta'}.
\lb{Tdef}
\en
By using \rlb{A1}, we find
\eq
\sfT=\pmat{\frac{1}{2}&\mu&0&0\\\mu&\frac{\mu^2}{2}&0&0\\0&0&-\frac{\mu}{2}&0\\0&0&0&-\frac{\mu}{2}},
\lb{T1}
\en
where the entries appear in the order (1,1), (2,2), (1,2), (2,1).
One readily finds that $\sfT$ has eigenvalues
\eq
\lambda_0=-\frac{\mu}{2},\quad
\lambda_\pm=\frac{\mu^2+1\pm\sqrt{\mu^4+14\mu^2+1}}{4},
\lb{lambda1}
\en
where $\lambda_0$ is doubly degenerate.
See Figure~\ref{f:mevs1}.

Let $\hO$ and $\hO'$ be arbitrary operators that act on a single spin, and denote by $\hO_j$ and $\hO'_k$ their copies at sites $j$ and $k$.
Then for any $1\le j<k\le L$, we see that the unnormalized correlation function has a representation
\eq
\sbkt{\Psi_\mu|\hO_j\hO'_k|\Psi_\mu}=\Tr[\sfD_{\hO}\,\sfT^{k-j-1}\sfD_{\hO'}\,\sfT^{L-k+j-1}],
\lb{OO}
\en
where  we defined the $4\times4$ matrix $\sfD_{\hO}$ corresponding to an arbitrary operator $\sfO$ by
\eq
(\sfD_{\hO})_{(\alpha,\beta),(\alpha',\beta')}=
\sum_{m,m'=-S}^S(\sfA^{(m)})_{\alpha,\alpha'}\bra{m}\hO\ket{m'}(\sfA^{(m')})_{\beta,\beta'}.
\en

In particular, we find from explicit computations that the spin-spin correlation functions for any $j\ne k$ are given by
\eq
\sbkt{\hS^{(\rz)}_j\hS^{(\rz)}_k}_\mu=\frac{4}{3}\Bigl(\frac{\lambda_-}{\lambda_+}\Bigr)^{|j-k|},
\en
and
\eq
\sbkt{\hS^{(\rx)}_j\hS^{(\rx)}_k}_\mu=\sbkt{\hS^{(\ry)}_j\hS^{(\ry)}_k}_\mu=
\frac{(\lambda_++\frac{1}{2})(\lambda_++\frac{\mu^2}{2})}{2(\lambda_+-\frac{\mu^2+1}{4})\lambda_+}
\Bigl(\frac{\lambda_0}{\lambda_+}\Bigr)^{|j-k|},
\en
where
\eq
\sbkt{\ \cdot\ }_\mu=
\lim_{L\up\infty}\frac{\sbkt{\Psi_\mu|\cdot|\Psi_\mu}}{\sbkt{\Psi_\mu|\Psi_\mu}}
\en
denotes the expectation value in the infinite volume limit of the asymmetric VBS state.
Noting that $\lambda_+>0$, $|\lambda_-/\lambda_+|<1$, and $|\lambda_0/\lambda_+|<1$ for any $\mu\in[0,1]$, we see that the spin-spin correlation functions in the asymmetric VBS state always exhibit exponential decay accompanied by antiferromagnetic oscillation.
We note that the correlation lengths are uniformly bounded in the whole range $\mu\in[0,1]$.

More generally, one can prove that any truncated correlation function decays exponentially.
\begin{theorem}
For arbitrary single-site operators $\hO$, $\hO'$, arbitrary $\mu\in[0,1]$, and arbitrary $j,k\ge1$, we have
\eq
\Bigl|\sbkt{\hO_j\hO'_k}_\mu-\sbkt{\hO_j}_\mu\sbkt{\hO'_k}_\mu\Bigr|\le C\,\Bigl|\frac{\lambda_0}{\lambda_+}\Bigr|^{|j-k|},
\en 
where $C$ is a constant that may depend only on the operators and $\mu$.
\end{theorem}
We conclude that the $S=1$ asymmetric VBS state \rlb{mu1} is disordered for all $\mu\in[0,1]$.

We note that $\ket{\Psi_\mu}$ exhibits nonzero string order (which reveals the hidden antiferromagnetic order  \cite{denNijsRommelse}) in the z-direction for any $\mu\in(0,1]$.\footnote{
The string order in the x or y-direction presents only for $\mu=1$.
}
But this fact has little physical significance since the model does not have the $\ZZ$ symmetry (except at $\mu=1$ or 0).
See, e.g., chapter~8 of \cite{TasakiBook} for details.

\subsection{The interpolating Hamiltonians}\label{s:iH1}
The existence of the asymmetric VBS states suggests that the AKLT model \rlb{HAKLT} and the trivial model \rlb{Htrivial} are continuously connected.
We shall prove this fact explicitly by constructing interpolating Hamiltonians.

\begin{theorem}\label{t:S=1Ham}
There is a one-parameter family $\hH_\mu$ (with $\mu\in[-1,1]$) of short-ranged translation-invariant Hamiltonians such that $\hH_\mu$ depends on $\mu$ continuously and that $\hH_1=\hH_\mathrm{AKLT}$ and $\hH_{-1}=\hH_\mathrm{trivial}$.
For $\mu\in[0,1]$, the asymmetric VBS state $\ket{\Psi_\mu}$ is the unique gapped ground state of $\hH_\mu$.
For $\mu\in[-1,0]$, the zero state $\ket{\Phi_\mathrm{zero}}$ is the unique gapped ground state of $\hH_\mu$.
\end{theorem}
The theorem implies that the AKLT Hamiltonian \rlb{HAKLT} and the trivial Hamiltonian \rlb{Htrivial} for the $S=1$ chain belong to the same phase in the large space of models $\calM$.
This fact was already proved by Bachman and Nachtergaele through explicit models \cite{BachmannNachtergaele2012,BachmannNachtergaele2014}  and by Ogata through a general theory \cite{Ogata1,Ogata2,Ogata3}.

\medskip\noindent
{\em Proof:}\/
The proof consists of several steps.
To begin with, we construct Hamiltonians by using projection operators as in the original approach to the AKLT model \cite{AKLT1,AKLT2,TasakiBook}.
Note that the restriction of the asymmetric VBS state $\ket{\Psi_\mu}$ on an arbitrary pair of adjacent sites $j$ and $j+1$ is formed by a combination of the states
\eq
(\Sy_j\otimes\Sy_{j+1})\Bigl(\ket{\sigma}_{\!(j,\ra)}\otimes\pmu{(j,\rb),(j+1,\ra)}\otimes\ket{\sigma'}_{\!(j+1,\rb)}\Bigr),
\en
with $\sigma,\sigma'=\up,\dn$.
By using \rlb{mu} and \rlb{S}, we rewrite these four states in terms of the standard basis states for $S=1$ as
\eqg
\kone_j\kzero_{j+1}-\mu\kzero_j\kone_{j+1},\quad \kone_j\kmone_{j+1}-\frac{\mu}{2}\kzero_j\kzero_{j+1},\nl
\tfrac{1}{2}\kzero_j\kzero_{j+1}-\mu\kmone_j\kone_{j+1},\quad\kzero_j\kmone_{j+1}-\mu\kmone_j\kzero_{j+1}.
\lb{four}
\eng
Let us concentrate on the nine-dimensional Hilbert space corresponding to the spins at $j$ and $j+1$.
We let $\hP^{(\mu)}_{j,j+1}$ be the projection operator onto the five-dimensional subspace that is orthogonal to the four-dimensional subspace spanned by the states \rlb{four}.
We then define our Hamiltonian as
\eq
\hH'_\mu=\sum_{j=1}^L\hP^{(\mu)}_{j,j+1},
\lb{Hmu1}
\en
for any $\mu\in[0,1]$.
Note that $\hH'_\mu$ is continuous in $\mu$.
It is easily checked that this Hamiltonian with $\mu=1$ is nothing but the AKLT Hamiltonian \rlb{HAKLT}.
Noting that $\hP^{(\mu)}_{j,j+1}\ket{\Psi_\mu}=0$ by definition, we see that $\ket{\Psi_\mu}$ is a ground state of $\hH'_\mu$ with the ground state energy $0$ for any $\mu\in[0,1]$.

We conjecture that the asymmetric VBS state $\ket{\Psi_\mu}$ is the unique gapped ground state of $\hH'_\mu$ for any $\mu\in[0,1]$.
If we could verify this conjecture, we could simply use $\hH'_\mu$ as the interpolating Hamiltonian.
Unfortunately, the conjecture can be justified only partially, as we shall see below.\footnote{
\label{fn:UGGS}
The uniqueness of the ground state can indeed be proved for any $\mu\in[0,1]$ by a straightforward extension of the method developed in \cite{KennedyLiebTasaki1988}.
(See also section~7.1.3 of \cite{TasakiBook}.)
But we do not use this general result here.
It is also likely that the existence of a gap for any $\mu\in[0,1]$ can be proved by extending the method of Knabe \cite{Knabe} (see also section~7.1.4 of \cite{TasakiBook}).}
First, we recall that the VBS state $\ket{\Psi_1}$ is proved to be the unique gapped ground state of the AKLT Hamiltonian $\hH'_1=\hH_{\rm AKLT}$ \cite{AKLT2,Knabe,TasakiBook}.
Secondly, it is easy to show that the zero state $\ket{\Psi_0}$ is the unique gapped ground state of $\hH'_0$.\footnote{
\label{fn:H0}
We first prove the uniqueness.
From \rlb{four}, we see that any ground state should locally consist of $\kone_j\kzero_{j+1}$, $ \kone_j\kmone_{j+1}$, $\kzero_j\kzero_{j+1}$, or $\kzero_j\kmone_{j+1}$.
The only global solution is the zero state if we impose the periodic boundary condition.

To prove the existence of a gap, consider a state of the form $\ket{\boldsymbol{\sigma}}=\bigotimes_{j=1}^L\ket{\sigma_j}_j$ with $\sigma_j=\pm1,0$, and $\sigma_k\ne0$ for at least one $k$.
Then there is at least one pair $j$, $j+1$ of neighboring sites such that the spin configuration $(\sigma_j,\sigma_{j+1})$ differs from the four configurations listed above.
This means $\bra{\boldsymbol{\sigma}}\hH_0\ket{\boldsymbol{\sigma}}\ge1$.
Since $\ket{\boldsymbol{\sigma}}$ is orthogonal to the ground state $\ket{\Phi_\mathrm{zero}}$, we see that there is a gap, which is equal to 1.
}
Thirdly, these results for $\mu=1$ and $\mu=0$ and the rigorous perturbation theories \cite{BHM,BH,NSY} guarantee that $\hH'_\mu$ has a unique gapped ground state when $\mu\in[0,1]$ is sufficiently small or sufficiently close to 1.
Thus we have proved that the asymmetric VBS state $\ket{\Psi_\mu}$ is the unique gapped ground state of $\hH'_\mu$ when $\mu$ is sufficiently close to 0 or 1.

Next, we prepare another family of interpolating Hamiltonians by making use of the general theory of injective matrix product states developed by Fannes, Nachtergaele, and Werner \cite{FannesNachtergaeleWerner1992,PerezGarciaVerstraete}.
The theory applies to the asymmetric VBS state with $\mu\in(0,1]$, but not to that with $\mu=0$.
It is shown that there exists a family of short-ranged Hamiltonians $\hH''_\mu$ that depends continuously on $\mu$ such that $\ket{\Psi_\mu}$ is a unique gapped ground state of $\hH''_\mu$ for any $\mu\in(0,1]$.
See Appendix~\ref{a:injective} for a technical point that is necessary for the use of the general theory.

We can now define desired interpolating Hamiltonians for $\mu\in[0,1]$.
Recall that we have not proved the existence of a gap in $\hH'_\mu$ for intermediate $\mu$ and that $\hH''_\mu$ is not defined for $\mu=0$.
We, therefore, combine the two interpolating Hamiltonians as
\eq
\hH_\mu=\hH'_\mu+f(\mu)\,\hH''_\mu,
\en
with an arbitrary continuous function $f(\mu)$ such that $f(0)=f(1)=0$ and $f(\mu)>0$ for $\mu\in(0,1)$.
Clearly $\hH_\mu$ is defined for all $\mu\in[0,1]$, depends continuously on $\mu$, coincides with $\hH_\mathrm{AKLT}$ when $\mu=1$, and is rigorously guaranteed to have $\ket{\Psi_\mu}$ as its unique gapped ground state.

We finally have to note that our $\hH_0$ and the trivial Hamiltonian \rlb{Htrivial} are very different, although they share the zero state \rlb{zero} as their ground states.
In order to connect $\hH_0$ and the trivial Hamiltonian \rlb{Htrivial}, we  extended the range of $\mu$ to $[-1,1]$ and define
\eq
\hH_\mu=-\mu\,\hH_\mathrm{trivial}+(1+\mu)\,\hH_0,
\lb{Hps}
\en
for $\mu\in[-1,0]$.
One can easily verify that, for $\mu\in[-1,0]$, the Hamiltonian $\hH_\mu$ always has the zero state \rlb{zero} as its unique gapped ground state.~\qedm

\subsection{The edge states and spin pumping}
Although the Hamiltonian \rlb{Hmu1} has a unique (and probably gapped) ground state\footnote{See footnote~\ref{fn:UGGS}.} for any $\mu\in[0,1]$, its counterpart with open boundary conditions, i.e.,
\eq
\hH_\mu^\mathrm{open}=\sum_{j=1}^{L-1}\hP^{(\mu)}_{j,j+1},
\lb{Hmu1open}
\en
has four-fold degenerate ground states given by
\eq
\ket{\Psi_\mu^{\sigma,\sigma'}}=\Bigl(\bigotimes_{j=1}^L\Sy_j\Bigr)\Bigl(\ket{\sigma}_{\!(1,\ra)}\otimes\Bigl(\bigotimes_{j=1}^{L-1}\pmu{(j,\rb),(j+1,\ra)}\Bigr)\otimes\ket{\sigma'}_{\!(L,\rb)}\Bigr),
\lb{mu1ss}
\en
for any $\sigma,\sigma'=\up,\dn$.

As is clear from the expression \rlb{mu1ss}, the degeneracy reflects the emergence of the $S=\frac{1}{2}$ degrees of freedom at the edges.
It should be noted that this is valid for the Hamiltonian \rlb{Hmu1} with any $\mu\in[0,1]$.
We see that there is something ``topological'' even in the Hamiltonian $\hH_0$, which has the trivial zero state \rlb{zero} as the ground state.
This interesting point was emphasized by Bachmann and Nachtergaele \cite{BachmannNachtergaele2014}.

By setting $\mu=0$ in \rlb{mu}, we see $\ket{\psi^{(0)}_{p,q}}=\uket_p\dket_q$, and hence $\ket{\Psi_0^{\dn,\up}}=\otimes_{j=1}^L\ket{0}_j$ from \rlb{mu1ss}.
This means that one can continuously deform the AKLT state $\ket{\Psi_1^{\dn,\up}}$ with apparent edge spins into the zero state $\otimes_{j=1}^L\ket{0}_j$ without edge spins.
One can further apply the inverse deformation, but with left and right reversed, to continuously modify the zero state $\otimes_{j=1}^L\ket{0}_j$ into another AKLT state $\ket{\Psi_1^{\up,\dn}}$.
Combining these two processes, we have a continuous path of states that connect the two AKLT states $\ket{\Psi_1^{\dn,\up}}$ and $\ket{\Psi_1^{\up,\dn}}$.
The two states are, of course, the ground states of the same Hamiltonian $\hH_1^\mathrm{open}$, but differ in their edge states.

The process that continuously brings $\ket{\Psi_1^{\dn,\up}}$ into $\ket{\Psi_1^{\up,\dn}}$ can be regarded as an example of spin pumping \cite{Thouless,Shindou,KunoHatsugai2021}.
Since the $S^{(\rz)}=\pm1/2$ spins at both ends are exchanged, one can say that $S^{(\rz)}=1$ is pumped from right to left in this process.
We note that, unlike in the previous examples \cite{Shindou,KunoHatsugai2021}, the pumping is realized here only by using translationally invariant Hamiltonians.
The existence of nonzero spin pumping in this process is not an accident but a consequence of the fact that the path connects the two states, $\ket{\Psi_1^{\dn,\up}}$ and $\ket{\Psi_1^{\up,\dn}}$, in the nontrivial SPT phase via the state $\otimes_{j=1}^L\ket{0}_j$  in the trivial SPT phase.
This observation is closely related to recent studies of the topological characterization of paths in the space of models \cite{Kapustin2020,Wen2021,Bachmann2022}.
See \cite{TasakiLiebConference} for a brief announcement, and \cite{TasakiNext} for details.

\subsection{The modification operator}\label{s:MO}
Let us discuss an interesting method of generating the asymmetric VBS state with $\mu>0$ by modifying the AKLT state.\footnote{\label{fn:Katsura1}
We learned this method from Hosho Katsura.}
Consider the chain $\{1,\ldots,L\}$ with open boundary conditions, and define the Hermitian operator
\eq
\hM_\mu=\mu^{\sum_{j=1}^Lj\hSz_j},
\lb{Mmu}
\en
which is invertible for $\mu\in(0,1]$.
By writing $\hM=\exp[-i\sum_{j=1}^L\kappa j\hSz_j]$ with $\kappa=i\log\mu$, we see that $\hM$ can be regarded as the twist operator of Bloch \cite{Bohm} and Lieb, Schultz, and Mattis \cite{LSM}, but with purely imaginary rotation angle.

If $\ket{\Phi}$ is a nearly translation-invariant state on the open chain, the state $\hM_\mu\ket{\Phi}$ with $\mu<1$ is, in general, a highly non-translation-invariant state in which configurations with large negative $\sum_{j=1}^Lj\hSz_j$ (which may be regarded as polarization) have large weights.\footnote{This may be interpreted as a non-Hermitian skin effect in a spin chain.}
But the situation is essentially different for the AKLT state.

To see how $\hM_\mu$ acts on the AKLT state $\ket{\Phi_{\rm AKLT}^{\sigma,\sigma'}}=\ket{\Psi_1^{\sigma,\sigma'}}$, we first note that
\eq
\hM_\mu\Bigl(\bigotimes_{j=1}^L\Sy_j\Bigr)=\Bigl(\bigotimes_{j=1}^L\Sy_j\Bigr)\mu^{\sum_{j=1}^Lj(\hSz_{j,\ra}+\hSz_{j,\rb})},
\lb{MS}
\en
and then observe that
\eqa
\mu^{j\hSz_{j,\rb}+(j+1)\hSz_{j+1,\ra}}\ket{\psi^{(1)}_{(j,\rb),(j+1,\ra)}}
&=\mu^{j\hSz_{j,\rb}+(j+1)\hSz_{j+1,\ra}}\bigl(\uket_{j,\rb}\dket_{j+1,\ra}-\dket_{j,\rb}\uket_{j+1,\ra}\bigr)
\nl&=\mu^{-1/2}\,\uket_{j,\rb}\dket_{j+1,\ra}-\mu^{1/2}\,\dket_{j,\rb}\uket_{j+1,\ra}
\nl&=\mu^{-1/2}\,\pmu{(j,\rb),(j+1,\ra)}.
\lb{Mmup}
\ena
We then find from \rlb{mu1ss} and \rlb{MS} that
\eq
\hM_\mu\ket{\Psi_1^{\sigma,\sigma'}}=\mu^{\sigma+L\sigma'}\mu^{-L/2}\,\ket{\Psi_\mu^{\sigma,\sigma'}},
\lb{426}
\en 
where we understand that $\sigma,\sigma'=\pm\frac{1}{2}$.
Remarkably, we get the asymmetric VBS state $\ket{\Psi_\mu^{\sigma,\sigma'}}$ by modifying the AKLT state $\ket{\Psi_1^{\sigma,\sigma'}}$ with the operator $\hM_\mu$.

Since the AKLT Hamiltonian $\hH_1^\mathrm{open}$ and its modification $\hK_\mu=\hM_\mu\hH_1^\mathrm{open}\hM_\mu^{-1}$ have exactly the same spectra, one readily finds that the asymmetric VBS states $\ket{\Psi_\mu^{\sigma,\sigma'}}$ are the only ``ground states'' accompanied by a nonzero gap of the ``Hamiltonian'' $\hK_\mu$.
Of course, $\hK_\mu$ is not what we want since it is not Hermitian.
The operator $\hK'_\mu=\hM_\mu^{-1}\hH_1^\mathrm{open}\hM_\mu^{-1}$ is Hermitian and has $\ket{\Psi_\mu^{\sigma,\sigma'}}$ as ground states, but it is extremely long-ranged.
By using the procedure known as Witten's conjugation \cite{Witten,WoutersKatsuraSchuricht}, one can modify the AKLT Hamiltonian $\hH_1^\mathrm{open}$ to get a Hermitian operator with short-ranged interactions, which we call $\hH_\mu^\mathrm{WC}$, that has $\ket{\Psi_\mu^{\sigma,\sigma'}}$ as ground states.
It turns out that our $\hH_\mu^\mathrm{open}$ is a particular case of $\hH_\mu^\mathrm{WC}$.
See Appendix~\ref{s:Witten}.

\section{The asymmetric VBS state for the $S=2$ chain}
\label{s:S=2}
We now consider a quantum spin system with $S=2$ on the same periodic chain with $L$ sites (where $L$ is an arbitrary positive integer) and define the asymmetric VBS state.
Again the asymmetric VBS state continuously interpolates between the AKLT state \rlb{AKLT2} and the zero state \rlb{zero}, but now always possesses the three types of symmetry, i.e., the time-reversal symmetry, the $\ZZ$ symmetry, and the bond-centered inversion symmetry.  
As noted in section~\ref{s:intro}, Pollmann, Turner, Berg, and Oshikawa \cite{PollmannTurnerBergOshikawa2012} constructed and discussed a family of matrix product states for the $S=2$ chain that plays the same role.
Our approach may be advantageous since it is based on the unified construction based on the asymmetric valence-bonds, and leads to much simpler matrix product states where the basic properties can be calculated without a computer.
We also note that the $S=2$ asymmetric VBS state is essentially the same as one of the general matrix product states defined in \cite{ASZ}.\footnote{\label{fn:Katsura2}
We learned this fact from Hosho Katsura.}

\subsection{The state and its symmetry}
The construction of the asymmetric VBS state for the $S=2$ chain is analogous to that for the $S=1$ chain.
We shall be brief and discuss only the main differences.

This time we represent a spin with $S=2$ as a composite of four spins with $S=\frac{1}{2}$.
To be precise, consider four $S=\frac{1}{2}$ spins labeled as a, b, c, and d.
Then we define the projection operator onto the $S=2$ states (or, equivalently, the symmetrization operator) by
\eqa
\Sy&=\ket{2}\,{}_\rd\!\ubra_\rc\!\ubra_\rb\!\ubra_\ra\!\ubra
\nl&+
\kone\frac{{}_\rd\!\dbra_\rc\!\ubra_\rb\!\ubra_\ra\!\ubra+{}_\rd\!\ubra_\rc\!\dbra_\rb\!\ubra_\ra\!\ubra+{}_\rd\!\ubra_\rc\!\ubra_\rb\!\dbra_\ra\!\ubra+{}_\rd\!\ubra_\rc\!\ubra_\rb\!\ubra_\ra\!\dbra}{2}
\nl&+
\kzero\frac{{}_\rd\!\dbra_\rc\!\dbra_\rb\!\ubra_\ra\!\ubra+{}_\rd\!\dbra_\rc\!\ubra_\rb\!\dbra_\ra\!\ubra
+{}_\rd\!\dbra_\rc\!\ubra_\rb\!\ubra_\ra\!\dbra
+{}_\rd\!\ubra_\rc\!\dbra_\rb\!\dbra_\ra\!\ubra+{}_\rd\!\ubra_\rc\!\dbra_\rb\!\ubra_\ra\!\dbra
+{}_\rd\!\ubra_\rc\!\ubra_\rb\!\dbra_\ra\!\dbra}{\sqrt{6}}
\nl&
+\kmone\frac{{}_\rd\!\dbra_\rc\!\dbra_\rb\!\dbra_\ra\!\ubra+{}_\rd\!\dbra_\rc\!\dbra_\rb\!\ubra_\ra\!\dbra+{}_\rd\!\dbra_\rc\!\ubra_\rb\!\dbra_\ra\!\dbra+{}_\rd\!\ubra_\rc\!\dbra_\rb\!\dbra_\ra\!\dbra}{2}
\nl&
+\ket{\!-\!2}\,{}_\rd\!\dbra_\rc\!\dbra_\rb\!\dbra_\ra\!\dbra,
\lb{S2}
\ena
where $\ket{m}$ with $m=0,\pm1,\pm2$ denotes the basis states of a spin with $S=2$.

\begin{figure}
\centerline{\epsfig{file=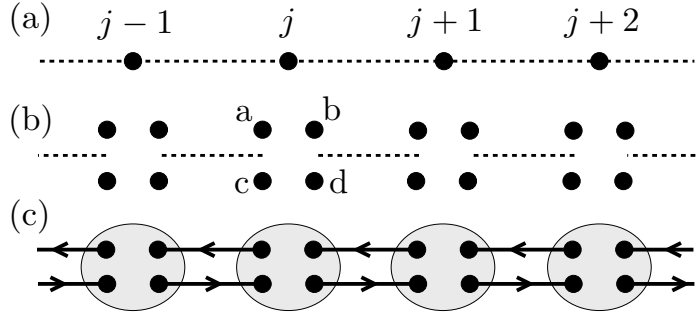,width=9truecm}}
\caption[dummy]{
The construction of the $S=2$ asymmetric VBS state.
(a)~The original chain with $L$ sites.
(b)~The quadruplicated chain.  We quadruplicate a single site $j$ into four sites $(j,\ra)$, $(j,\rb)$, $(j,\rc)$, and $(j,\rd)$.
(c)~The $S=2$ asymmetric VBS state.  
A gray oval denotes symmetrization of four spin $\frac{1}{2}$'s, which gives an $S=2$.
Note that the two asymmetric valence-bonds connecting two neighboring sites (in the original chain) have opposite orientations.
In this manner, the asymmetry is canceled, and one gets a state which is invariant under the time-reversal, the $\ZZ$, and the bond-centered inversion transformation.
}
\label{f:mVBS2}
\end{figure}

We define the quadruplicated chain with $4L$ sites that has four sites $(j,\ra)$, $(j,\rb)$, $(j,\rc)$, and $(j,\rd)$ corresponding to a single site $j$ in the original chain.
As before, we associate each site in the quadruplicated chain with a $S=\frac{1}{2}$ spin, and define the asymmetric VBS state by
\eq
\ket{\Psi_\mu}=\Bigl(\bigotimes_{j=1}^L\Sy_j\Bigr)\Bigl(\bigotimes_{j=1}^L\pmu{(j,\rb),(j+1,\ra)}\otimes\pmu{(j+1,\rc),(j,\rd)}\Bigr),
\lb{mu2}
\en
where, again, $\Sy_j$ is the operator \rlb{S2} for site $j$.
It is crucial here that the two asymmetric valence-bonds connecting the neighboring sites $j$ and $j+1$ have opposite orientations.
See Figure~\ref{f:mVBS2}.

It is again obvious that \rlb{mu2} with $\mu=1$ is the AKLT state \rlb{AKLT2}.
We also see, as in \rlb{P0}, that \rlb{mu2} with $\mu=0$ is proportional to the zero state \rlb{zero}.
We have confirmed that the asymmetric VBS state $\ket{\Psi_\mu}$ continuously interpolates between the AKLT state and the zero state for the $S=2$ chain.

Let us examine the symmetry of the $S=2$ asymmetric VBS state \rlb{mu2}.
By the bond-centered inversion $j\to L+1-j$, the state on the bond (L,1) transforms as
\eq
\pmu{(L,\rb),(1,\ra)}\otimes\pmu{(1,\rc),(L,\rd)}\to\pmu{(1,\ra),(L,\rb)}\otimes\pmu{(L,\rd),(1,\rc)}.
\en
Note that the resulting state is the same as the original if we switch the roles of the labels as $\ra\leftrightarrow\rc$ and $\rb\leftrightarrow\rd$.
We thus see, along with similar considerations for other bonds, that the asymmetric VBS state \rlb{mu2} as a whole is invariant under the bond-centered inversion.

Likewise, we see that the state on a bond $(j,j+1)$ transforms under the time-reversal \rlb{Tp} as
\eq
\hT\pmu{(j,\rb),(j+1,\ra)}\otimes\pmu{(j+1,\rc),(j,\rd)}=\pmu{(j+1,\ra),(j,\rb)}\otimes\pmu{(j,\rd),(j+1,\rc)},
\en
and under $\ZZ$ transformation as
\eqg
\hUy\pmu{(j,\rb),(j+1,\ra)}\otimes\pmu{(j+1,\rc),(j,\rd)}=\pmu{(j+1,\ra),(j,\rb)}\otimes\pmu{(j,\rd),(j+1,\rc)},\\
\hUz\pmu{(j,\rb),(j+1,\ra)}\otimes\pmu{(j+1,\rc),(j,\rd)}=\pmu{(j,\rb),(j+1,\ra)}\otimes\pmu{(j+1,\rc),(j,\rd)}.
\eng
We thus find that the asymmetric VBS state $\ket{\Psi_\mu}$ is invariant under both the time-reversal and the $\ZZ$ transformation.
The essential difference from the $S=1$ state comes from the fact that the asymmetry (or the direction) on each pair of neighboring sites can be canceled in the $S=2$ state, where one has two asymmetric valence-bonds connecting neighboring sites.

\subsection{The matrix product representation and correlation functions}
\begin{figure}
\centerline{\epsfig{file=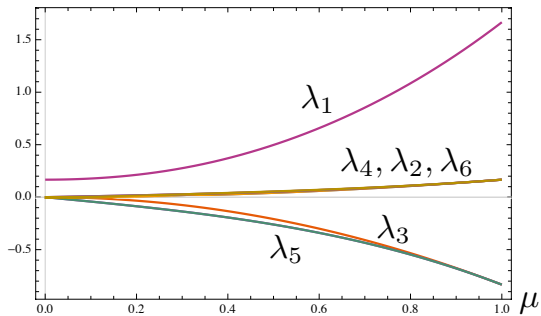,width=7truecm}}
\caption[dummy]{
The eigenvalues \rlb{lambda2}.
They are ordered as $\lambda_1>\lambda_4>\lambda_2>\lambda_6>0>\lambda_3>\lambda_5$ for $\mu\in(0,1)$.
One can hardly distinguish between $\lambda_4$, $\lambda_2$, and $\lambda_6$ in the figure.
}
\label{f:mevs2}
\end{figure}

The property of the $S=2$ asymmetric VBS state becomes clear by representing it as a matrix product state.
In this case, we have three alternatives for each bond, namely, 
$-\mu\uket_{\!(j,\rb)}\uket_{\!(j,\rd)}\dket_{\!(j+1,\ra)}\dket_{\!(j+1,\rc)}$
or
$\uket_{\!(j,\rb)}\dket_{\!(j,\rd)}\dket_{\!(j+1,\ra)}\uket_{\!(j+1,\rc)}+\mu^2\dket_{\!(j,\rb)}\uket_{\!(j,\rd)}\uket_{\!(j+1,\ra)}\dket_{\!(j+1,\rc)}$ or
$-\mu\dket_{\!(j,\rb)}\dket_{\!(j,\rd)}\uket_{\!(j+1,\ra)}\uket_{\!(j+1,\rc)}$.
This leads to the representation \rlb{mps}, where we now set $S=2$, with five $3\times 3$ matrices
\eqg
\sfA^{(2)}=\pmat{0&0&0\\0&0&0\\-\mu&0&0},\quad
\sfA^{(1)}=\frac{1}{2}\pmat{0&0&0\\\sqrt{\mu(1+\mu^2)}&0&0\\0&-\sqrt{\mu(1+\mu^2)}&0},\quad
\sfA^{(0)}=\frac{1}{\sqrt{6}}\pmat{-\mu&0&0\\0&1+\mu^2&0\\0&0&-\mu},\nl
\sfA^{(-1)}=\frac{1}{2}\pmat{0&-\sqrt{\mu(1+\mu^2)}&0\\0&0&\sqrt{\mu(1+\mu^2)}\\0&0&0},\quad
\sfA^{(-2)}=\pmat{0&0&-\mu\\0&0&0\\0&0&0}.
\lb{A2}
\eng
The corresponding transfer matrix \rlb{Tdef} is given by
\eq
\sfT=\pmat{
\frac{\mu^2}{6}&\frac{\mu(1+\mu^2)}{4}&\mu^2&0&0&0&0&0&0\\
\frac{\mu(1+\mu^2)}{4}&\frac{(1+\mu^2)^2}{6}&\frac{\mu(1+\mu^2)}{4}&0&0&0&0&0&0\\
\mu^2&\frac{\mu(1+\mu^2)}{4}&\frac{\mu^2}{6}&0&0&0&0&0&0\\
0&0&0&-\frac{\mu(1+\mu^2)}{6}&-\frac{\mu(1+\mu^2)}{4}&0&0&0&0\\
0&0&0&-\frac{\mu(1+\mu^2)}{4}&-\frac{\mu(1+\mu^2)}{6}&0&0&0&0\\
0&0&0&0&0&-\frac{\mu(1+\mu^2)}{6}&-\frac{\mu(1+\mu^2)}{4}&0&0\\
0&0&0&0&0&-\frac{\mu(1+\mu^2)}{4}&-\frac{\mu(1+\mu^2)}{6}&0&0\\
0&0&0&0&0&0&0&\frac{\mu^2}{6}&0\\
0&0&0&0&0&0&0&0&\frac{\mu^2}{6}
},
\lb{T2}
\en
where the entries appear in the order (1,1), (2,2), (3,3), (1,2), (2,3), (2,1), (3,2), (1,3), (3,1).
The eigenvalues of $\sfT$ are easily computed (even without a computer) as
\eqg
\lambda_1=\tfrac{1}{12}\{1+9\mu^2+\mu^4+\sqrt{1+8\mu^2+63\mu^4+8\mu^6+\mu^8}\},\nl
\lambda_2=\tfrac{1}{12}\{1+9\mu^2+\mu^4-\sqrt{1+8\mu^2+63\mu^4+8\mu^6+\mu^8}\},\quad\lambda_3=-5\mu^2/6,\nl
\lambda_4=\tfrac{1}{12}\mu(1+\mu^2),\quad\lambda_5=-\tfrac{5}{12}\mu(1+\mu^2),\quad\lambda_6=\mu^2/6,
\lb{lambda2}
\eng
where $\lambda_1$, $\lambda_2$, and $\lambda_3$ are nondegenerate and $\lambda_4$, $\lambda_5$, and $\lambda_6$ are doubly degenerate.
See Figure~\ref{f:mevs2}.
It is found that $\lambda_1>0$ and $|\lambda_j/\lambda_1|<1$ for any $j=2,\ldots,6$ and $\mu\in[0,1]$.
Here we do not present explicit calculations of expectation values, and state the following general result that follows from the above observations.
\begin{theorem}
For arbitrary single-site operators $\hO$, $\hO'$, arbitrary $\mu\in[0,1]$, and arbitrary $j,k\ge1$, we have
\eq
\Bigl|\sbkt{\hO_j\hO'_k}_\mu-\sbkt{\hO_j}_\mu\sbkt{\hO'_k}_\mu\Bigr|\le C\,\Bigl|\frac{\lambda_5}{\lambda_1}\Bigr|^{|j-k|},
\en 
where $C$ is a constant that may depend only on the operators and $\mu$.
\end{theorem}
We conclude that the $S=2$ asymmetric VBS state \rlb{mu2} is disordered for all $\mu\in[0,1]$.
We also note that it exhibits no hidden antiferromagnetic order.

\subsection{The interpolating Hamiltonians}\label{s:iH2}
As in the $S=1$ chain, we can construct a family of Hamiltonians that interpolates between the AKLT model \rlb{HAKLT} and the trivial model \rlb{Htrivial}.
We have the following theorem, which is identical to Theorem~\ref{t:S=1Ham} for $S=1$ except for the final statement about the symmetry.

\begin{theorem}
There is a one-parameter family $\hH_\mu$ (with $\mu\in[-1,1]$) of short-ranged translation-invariant Hamiltonians such that $\hH_\mu$ depends on $\mu$ continuously and that $\hH_1=\hH_\mathrm{AKLT}$ and $\hH_{-1}=\hH_\mathrm{trivial}$.
For $\mu\in[0,1]$, the asymmetric VBS state $\ket{\Psi_\mu}$ is the unique gapped ground state of $\hH_\mu$.
For $\mu\in[-1,0]$, the zero state $\ket{\Phi_\mathrm{zero}}$ is the unique gapped ground state of $\hH_\mu$.
Moreover, $\hH_\mu$ is invariant under the time-reversal, the $\ZZ$, and the bond-centered inversion transformations for any $\mu\in[-1,1]$.
\end{theorem}
The theorem establishes that, for the $S=2$ spin chain, the AKLT Hamiltonian \rlb{HAKLT} and the trivial Hamiltonian \rlb{Htrivial} are continuously connected not only within the large space $\calM$ but also within the smaller spaces $\calM_\Sigma$ with symmetry, where $\Sigma$ can be either the time-reversal, the $\ZZ$, or the bond-centered inversion symmetry.\footnote{We note that no spin pumping takes place along this path.}
This provides us with a simple direct demonstration that the $S=2$ AKLT model belongs to the trivial SPT phase.

\medskip\noindent
{\em Proof:}\/
The proof is almost parallel to that of Theorem~\ref{t:S=1Ham}.
We shall be brief and discuss only some complications unique to the $S=2$ model.

As in the $S=1$ case, we first consider the projection operator $\hP^{(\mu)}_{j,j+1}$ onto the sixteen-dimensional subspace orthogonal to any of the nine states 
\eq
(\Sy_j\otimes\Sy_{j+1})\Bigl(\ket{\sigma}_{\!(j,\ra)}\otimes\ket{\sigma'}_{\!(j,\rc)}\otimes
\pmu{(j,\rb),(j+1,\ra)}\otimes\pmu{(j+1,\rc),(j,\rd)}
\otimes
\ket{\sigma''}_{\!(j+1,\rb)}\otimes\ket{\sigma'''}_{\!(j+1,\rd)}\Bigr),
\lb{nine}
\en
with $\sigma+\sigma'=0,\pm1$ and $\sigma''+\sigma'''=0,\pm1$.
If we define the Hamiltonian $\hH'_\mu$ as in \rlb{Hmu1}, we have $\hH'_1=\hH_\mathrm{AKLT}$.
One can again extend the method in \cite{KennedyLiebTasaki1988,TasakiBook} to prove that the $S=2$ asymmetric VBS state $\ket{\Psi_\mu}$ is the unique ground state of $\hH'_\mu$, but only for $\mu\in(0,1]$.
One easily finds by inspection that $\hH'_0$ has highly degenerate ground states.
This means that \rlb{Hmu1} is not a proper definition in this case.

To define interpolating Hamiltonians that work for $\mu=0$ as well, we consider the set of nine states as in \rlb{nine} defined on the three adjacent sites $j$, $j+1$, and $j+2$, and let $\hQ_{j,j+1,j+2}^{(\mu)}$ be the projection operator onto the 116-dimensional subspace orthogonal to all these states.
We then define the Hamiltonian $\hH'_\mu$ by
\eq
\hH'_\mu=\sum_{j=1}^L\hP^{(\mu)}_{j,j+1}+g(\mu)\,\hQ_{j,j+1,j+2}^{(\mu)},
\en
with a continuous function $g(\mu)$ such that $g(1)=0$ and $g(\mu)>0$ for any $\mu\in[0,1)$.
Now one can easily show, as in footnote~\ref{fn:H0}, that $\hH'_0$ has $\ket{\Psi_0}$ as its unique gapped ground state.
Then the rest of the proof is the same as in the case with $S=1$.
See again Appendix~\ref{a:injective} for the use of the general theory of injective matrix product states.~\qedm

\section{Discussion}
\label{s:S}
For spin chains with $S=1$ and $S=2$, we defined the asymmetric VBS states $\ket{\Psi_\mu}$ parametrized by $\mu\in[0,1]$.
The state $\ket{\Psi_\mu}$ continuously interpolates between the AKLT state $\ket{\Psi_1}=\ket{\Phi_\mathrm{AKLT}}$ and the zero state $\ket{\Psi_0}=\ket{\Phi_\mathrm{zero}}$, which is the ground state of the trivial Hamiltonian $\hH_\mathrm{trivial}$.
It is found that $\ket{\Psi_\mu}$ is disordered for any $\mu\in[0,1]$ in the sense that truncated correlation functions always decay exponentially.
We also constructed families of Hamiltonians $\hH_\mu$ (with $\mu\in[-1,1]$) interpolating between $\hH_\mathrm{AKLT}$ and $\hH_\mathrm{trivial}$ whose unique gapped ground state is the asymmetric VBS state $\ket{\Psi_\mu}$.
In short, the asymmetric VBS states are disordered ground states that continuously interpolate the disordered ground states of the AKLT model \rlb{HAKLT} and the trivial model \rlb{Htrivial}.

All these properties are common for the states with $S=1$ and 2, but there is an essential difference between the two cases.
The asymmetric VBS state with $S=2$ is invariant under the time-reversal, the $\ZZ$, and the bond-centered inversion transformations for any  $\mu\in[0,1]$, while that for $S=1$ has these symmetries only for $\mu=0$ and 1.
This is consistent with the fact that the AKLT model and the trivial model belong to the same SPT phase for $S=2$ but to different SPT phases for $S=1$.

Of course, the absence of symmetries in the asymmetric VBS state with $S=1$ does not show that the AKLT model and the trivial model are in different phases.
(This fact has been proved as a consequence of Ogata's general theory of SPT phases \cite{OgataZ2,Ogatainv,OgataCDM,HalSPT}.)
The role of these interpolating states is to explicitly demonstrate that the two models are continuously connected in the larger space $\calM$ of models with no restrictions on symmetry.
On the other hand, the fact that the $S=2$ asymmetric VBS state always possesses the three symmetries provides a simple direct proof that the AKLT model \rlb{HAKLT} and the trivial model \rlb{Htrivial} belong to the same SPT phase for the $S=2$ chain.

As was noted before, such families of interpolating disordered states were already constructed by Bachman and Nachtergale for $S=1$ \cite{BachmannNachtergaele2012,BachmannNachtergaele2014} and by Pollmann, Turner, Berg, and Oshikawa for $S=2$ \cite{PollmannTurnerBergOshikawa2012}.\footnote{Bachman and Nachtergale also constructed corresponding Hamiltonians \cite{BachmannNachtergaele2012,BachmannNachtergaele2014}.}
We here provided a unified construction that is also easy to understand intuitively.
After introducing the notion of asymmetric valence-bond, we only need to follow standard ideas and methods developed for the AKLT model.

\begin{figure}
\centerline{\epsfig{file=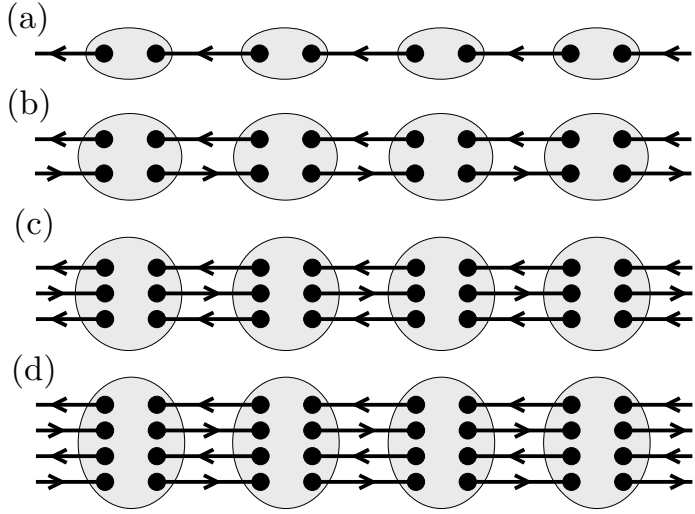,width=9truecm}}
\caption[dummy]{
The asymmetric VBS state with (a)~$S=1$, (b)~$S=2$, (c)~$S=3$, and (d)~$S=4$.
The directions of the arrows connecting two neighboring sites can be canceled for (b) and (d), but not for (a) and (c).
The cancelation implies that the AKLT state and the zero state can be continuously connected through disordered states that preserve all three types of symmetry relevant to SPT phases.
This simple observation provides us with an intuitive explanation of the qualitative difference between spin chains with even $S$ and odd $S$.
}
\label{f:mVBS3}
\end{figure}

We stress in particular that the symmetry properties of the asymmetric VBS states are manifest from their diagrammatic representations.
As we discussed in section~\ref{s:mu}, a single asymmetric valence-bond is not invariant under the time-reversal, the $\ZZ$, or the bond-centered inversion transformation.
The lack of symmetries is represented by the arrow in Figure~\ref{f:mVB}.
Then it is diagrammatically apparent that the $S=1$ asymmetric VBS state also lacks symmetries because the state has one arrow connecting two neighboring sites as in Figure~\ref{f:mVBS3}~(a).
In the $S=2$ asymmetric VBS state, on the other hand, there are two arrows with opposite directions connecting two sites as in Figure~\ref{f:mVBS3}~(b).
The fact that the state has all the three relevant symmetries is diagrammatically expressed as the cancelation of the directions of the two arrows.

This discussion naturally extends to spin chains with arbitrary integer spin.
If one can construct an asymmetric VBS state in which the directions of the arrows connecting two neighboring sites cancel, then the state has the three symmetries relevant to SPT phases.
In a spin chain with odd $S$, where one has an odd number of asymmetric valence-bonds connecting two sites, a cancelation is never possible.
See Figure~\ref{f:mVBS3}~(c).
Therefore the corresponding asymmetric VBS state cannot have the relevant symmetries.  
This is consistent with the fact that the odd $S$ AKLT model belongs to a nontrivial SPT phase.  
In a spin chain with even $S$, where one has an even number of asymmetric valence-bonds connecting two sites, one can arrange asymmetric valence-bonds so that the directions cancel.  
See Figure~\ref{f:mVBS3}~(d).
This means that the AKLT state and the trivial state are continuously connected through disordered (ground) states that have the time-reversal, the $\ZZ$, and the bond-centered inversion symmetries.  
This (along with the construction of interpolating Hamiltonians) shows that the AKLT model \rlb{HAKLT} belongs to the trivial SPT phase in a spin chain with even $S$.
In this manner, one can intuitively understand the qualitative difference between spin chains with even $S$ and odd $S$, discovered by Oshikawa in 1992 \cite{Oshikawa92} and finally explained by Pollmann, Turner, Berg, and Oshikawa in 2009 \cite{PollmannTurnerBergOshikawa2012}, by means of diagrams of asymmetric valence-bonds.

\appendix
\section{Injective matrix product states}
\label{a:injective}
In sections~\ref{s:iH1} and \ref{s:iH2}, we constructed Hamiltonians based on the general theory of injective matrix product states developed by Fannes, Nachtergaele, and Werner \cite{FannesNachtergaeleWerner1992,PerezGarciaVerstraete}.
To apply the general theory, however, the set of matrices representing a quantum state as in \rlb{mps} should be injective in the sense that (i)~the corresponding transfer matrix \rlb{Tdef} has a positive nondegenerate eigenvalue $\lamax$ such that any other eigenvalue $\lambda$ satisfies $|\lambda|<\lamax$, and (ii)~the set of matrices satisfy the normalization condition
\eq
\sum_{m=-S}^S\sfA^{(m)}(\sfA^{(m)})^\dagger=\lamax\,\sfI.
\lb{injnorm}
\en

From \rlb{lambda1} and \rlb{lambda2} one sees that the sets of matrices \rlb{A1} and \rlb{A2} for the asymmetric VBS states with $S=1$ and 2 satisfy the condition (i).
But one also finds that the normalization condition \rlb{injnorm} is not satisfied unless $\mu=1$ (where the state is nothing but the AKLT state).
By following a general procedure (see, e.g., \cite{Ogata1}), however, one can define sets of matrices that represent the same quantum states and also satisfy the conditions (i) and (ii), provided that $\mu>0$.

Let $(v_{\alpha,\beta})_{\alpha,\beta=1,\ldots,D}$ (where $D=2$ for $S=1$ and $D=3$ for $S=2$) be the eigenvector of the transfer matrix $\sfT$ corresponding to the eigenvalue $\lamax$, i.e. 
\eq
\sum_{\alpha',\beta'=1}^D(\sfT)_{(\alpha,\beta),(\alpha',\beta')}\,v_{\alpha',\beta'}=\lamax\,v_{\alpha,\beta}.
\lb{eveq}
\en
From \rlb{T1} and \rlb{T2} we see that the eigenvector has the form $v_{\alpha,\beta}=u_\alpha\,\delta_{\alpha,\beta}$, where $u_\alpha$ can be chosen to satisfy $u_\alpha>0$ for any $\alpha$ provided that $\mu>0$.
Substituting \rlb{Tdef} into \rlb{eveq} we find
\eq
\sum_m\sum_{\alpha'}(\sfA^{(m)})_{\alpha,\alpha'}(\sfA^{(m)})_{\beta,\alpha'}\,u_{\alpha'}=\lamax\,u_\alpha\,\delta_{\alpha,\beta}.
\lb{AAA}
\en
We define  new matrices $\sfB^{(m)}$ by
\eq
(\sfB^{(m)})_{\alpha,\alpha'}=\frac{1}{\sqrt{u_\alpha}}\,(\sfA^{(m)})_{\alpha,\alpha'}\,\sqrt{u_{\alpha'}}.
\lb{BBB}
\en
Clearly $\sfB^{(m)}$ defines the same quantum state \rlb{mps} as the original $\sfA^{(m)}$.
It is also easy to see that the transfer matrix defined from $\sfB^{(m)}$ as in \rlb{Tdef} has exactly the same eigenvalues as the original $\sfT$.
Finally we see from \rlb{AAA} and \rlb{BBB} that
\eqa
\Bigl(\sum_m\sfB^{(m)}(\sfB^{(m)})^\dagger\Bigr)_{\alpha,\beta}&=\sum_m\sum_{\alpha'}(\sfB^{(m)})_{\alpha,\alpha'}(\sfB^{(m)})_{\beta,\alpha'}
=\sum_m\sum_{\alpha'}\frac{1}{\sqrt{u_\alpha u_\beta}}\,(\sfA^{(m)})_{\alpha,\alpha'}(\sfA^{(m)})_{\beta,\alpha'}\,u_{\alpha'}
\nl&=\frac{1}{\sqrt{u_\alpha u_\beta}}\,\lamax\,u_\alpha\,\delta_{\alpha,\beta}=\lamax\,\delta_{\alpha,\beta},
\ena
which is the desired normalization condition \rlb{injnorm}.

\section{Witten's conjugation}
\label{s:Witten}
Here we follow \cite{WoutersKatsuraSchuricht} and discuss how one can modify the $S=1$ AKLT Hamiltonian by using Witten's conjugation \cite{Witten}.
See, in particular, section~3.4 of \cite{WoutersKatsuraSchuricht}, where the $q$-deformed AKLT state is treated in a similar manner.\footnote{\label{fn:Katsura3}
The material in this appendix is primarily due to Hosho Katsura.}

As in section~\ref{s:MO}, we consider an $S=1$ spin system on the open chain $\{1,\ldots,L\}$.
The AKLT Hamiltonian \rlb{HAKLT} for the open chain, which is denoted as $\hH^\mathrm{open}_1$ in section~\ref{s:S=1}, is
\eq
\hH^\mathrm{open}_\mathrm{AKLT}=\sum_{j=1}^{L-1}\hP_{j,j+1}^{S_\mathrm{tot}=2}.
\lb{HAKLTO}
\en
The mutually orthogonal normalized states on sites $j$ and $j+1$ that have $S_\mathrm{tot}=2$ are given by
\eqg
\ket{\xi^{(2)}_{j,j+1}}=\kone_j\kone_{j+1},\quad 
\ket{\xi^{(1)}_{j,j+1}}=\tfrac{1}{\sqrt{2}}(\kone_j\kzero_{j+1}+\kzero_j\kone_{j+1}),\quad
\nl
\ket{\xi^{(0)}_{j,j+1}}=\tfrac{1}{\sqrt{6}}(\kone_j\kmone_{j+1}+2\kzero_j\kzero_{j+1}+\kmone_j\kone_{j+1}), \nl
\ket{\xi^{(-1)}_{j,j+1}}=\tfrac{1}{\sqrt{2}}(\kzero_j\kmone_{j+1}+\kmone_j\kzero_{j+1}),\quad
\ket{\xi^{(-2)}_{j,j+1}}=\kmone_j\kmone_{j+1}.
\eng
Let us rewrite the projection operator as
\eq
\hP_{j,j+1}^{S_\mathrm{tot}=2}=\sum_{m=-2}^2\ket{\xi^{(m)}_{j,j+1}}\bra{\xi^{(m)}_{j,j+1}}=\sum_{m=-2}^2(\hA^{(m)}_{j,j+1})^\dagger\hA^{(m)}_{j,j+1},
\lb{PAA}
\en
with $\hA^{(m)}_{j,j+1}=\ket{\xi^{(0)}_{j,j+1}}\bra{\xi^{(m)}_{j,j+1}}$.
From \rlb{HAKLTO} and \rlb{PAA}, one finds that a state $\ket{\Phi}$ is a ground state of $\hH^\mathrm{open}_\mathrm{AKLT}$ if and only if $\hA^{(m)}_{j,j+1}\ket{\Phi}=0$ for all $m=-2,\ldots,2$ and $j=1,\ldots,L-1$.

We now take Witten's conjugation of $\hA^{(m)}_{j,j+1}$ and define
\eq
\hB^{(m)}_{\mu;j,j+1}=\alpha^{(m)}_{\mu;j,j+1}\,\hM_\mu\,\hA^{(m)}_{j,j+1}\,\hM_\mu^{-1},
\en
where $\alpha^{(m)}_{\mu;j,j+1}\ne0$ is an arbitrary constant.
Although $\hM_\mu$ defined by \rlb{Mmu} is highly nonlocal, $\hB^{(m)}_{\mu;j,j+1}$ is a local operator that acts nontrivially only on sites $j$ nd $j+1$.
Clearly a state $\ket{\Phi}$ is a ground state of $\hH^\mathrm{open}_\mathrm{AKLT}$ if and only if $\hB^{(m)}_{\mu;j,j+1}\hM_\mu\ket{\Phi}=0$ for all $m$ and $j$.
We define the new Hamiltonian, which is of course Hermitian, by
\eq
\hH^\mathrm{WC}_\mu=\sum_{j=1}^{L-1}\sum_{m=-2}^2(\hB^{(m)}_{\mu;j,j+1})^\dagger\hB^{(m)}_{\mu;j,j+1}.
\en
Recalling that $\ket{\Psi^{\sigma,\sigma'}_1}$ (see \rlb{mu1ss}) are the only ground state of $\hH^\mathrm{open}_\mathrm{AKLT}$ and that  $\hM_\mu\ket{\Psi^{\sigma,\sigma'}_1}$ is proportional to $\ket{\Psi^{\sigma,\sigma'}_\mu}$ as in \rlb{426}, we see that the four asymmetric VBS state $\ket{\Psi^{\sigma,\sigma'}_\mu}$ (with $\sigma,\sigma'=\up,\dn$) are the only ground states of $\hH^\mathrm{WC}_\mu$.

There is great freedom in the choice of $\hH^\mathrm{WC}_\mu$ since the constants $\alpha^{(m)}_{\mu;j,j+1}\ne0$ are arbitrary.
We shall discuss a natural choice of the constants.
Noting that 
\eq
(\hB^{(m)}_{\mu;j,j+1})^\dagger\hB^{(m)}_{\mu;j,j+1}=|\alpha^{(m)}_{\mu;j,j+1}|^2\,\mu^{-j\hSz_j-(j+1)\hSz_{j+1}}\ket{\xi^{(m)}_{j,j+1}}\bra{\xi^{(m)}_{j,j+1}}\,\mu^{-j\hSz_j-(j+1)\hSz_{j+1}},
\en 
we can choose the constant $\alpha^{(m)}_{\mu;j,j+1}$ so that the right-hand side is written as a projection $\ket{\tilde{\xi}^{(m)}_{\mu;j,j+1}}\bra{\tilde{\xi}^{(m)}_{\mu;j,j+1}}$ where $\ket{\tilde{\xi}^{(m)}_{\mu;j,j+1}}=\mathrm{(const)}\mu^{-j\hSz_j-(j+1)\hSz_{j+1}}\ket{\xi^{(m)}_{j,j+1}}$ is normalized.
Explicit calculation shows that
\eqg
\ket{\tilde{\xi}^{(2)}_{\mu;j,j+1}}=\kone_j\kone_{j+1},\quad 
\ket{\tilde{\xi}^{(1)}_{\mu;j,j+1}}=\tfrac{1}{\sqrt{1+\mu^2}}(\mu\kone_j\kzero_{j+1}+\kzero_j\kone_{j+1}),\quad
\nl
\ket{\tilde{\xi}^{(0)}_{\mu;j,j+1}}=\tfrac{1}{\sqrt{4+\mu^2+\mu^{-2}}}(\mu\kone_j\kmone_{j+1}+2\kzero_j\kzero_{j+1}+\mu^{-1}\kmone_j\kone_{j+1}), \nl
\ket{\tilde{\xi}^{(-1)}_{\mu;j,j+1}}=\tfrac{1}{\sqrt{1+\mu^2}}(\mu\kzero_j\kmone_{j+1}+\kmone_j\kzero_{j+1}),\quad
\ket{\tilde{\xi}^{(-2)}_{\mu;j,j+1}}=\kmone_j\kmone_{j+1}.
\eng
Since it is easily checked that these states are orthogonal to the four states \rlb{four}, we see that the Hamiltonian $\hH^\mathrm{WC}_\mu$ with this choice of constants is identical to our Hamiltonian $\hH_\mu^\mathrm{open}$ in \rlb{Hmu1open}.

\bigskip
{\small
It is our pleasure to thank Hosho Katsura for his valuable discussion and indispensable comments and for allowing us to include some of his contributions (see footnotes~\ref{fn:Katsura1}, \ref{fn:Katsura2}, and \ref{fn:Katsura3}) in the present paper, and 
Sven Bachmann,
Yasuhiro Hatsugai,
Marius Lemm,
Bruno Nachtergaele,
Yoshiko Ogata,
and
Masaki Oshikawa
for useful discussions.
H.T. was supported by JSPS Grants-in-Aid for Scientific Research No. 22K03474.
}


\begin{thebibliography}{99}

\bibitem{Haldane1981}
F.D.M. Haldane, 
{\em Ground State Properties of Antiferromagnetic Chains with Unrestricted Spin: Integer Spin Chains as Realisations of the $O(3)$ Non-Linear Sigma Model}\/,
ILL preprint SP-81/95 (1981).
\\\url{https://arxiv.org/abs/1612.00076}

\bibitem{Haldane1983a}
F.D.M. Haldane, 
{\em Continuum dynamics of the 1-D Heisenberg antiferromagnet: identification with the $O(3)$ nonlinear sigma model}\/,
Phys. Lett. {\bf 93A}, 464--468 (1983).
\\\url{http://www.sciencedirect.com/science/article/pii/037596018390631X}

\bibitem{Haldane1983b}
F.D.M. Haldane, 
{\em Nonlinear field theory of large-spin Heisenberg antiferromagnets: semiclassically quantized solitons of the one-dimensional easy-axis N\'eel state}\/,
Phys. Rev. Lett. {\bf 50} 1153--1156 (1983).
\\\url{https://journals.aps.org/prl/abstract/10.1103/PhysRevLett.50.1153}

\bibitem{TasakiBook}
H. Tasaki,
{\em Physics and mathematics of quantum many-body systems}\/, Graduate Texts in Physics (Springer, 2020).

\bibitem{HastingsKoma}
M.B. Hastings and T. Koma,
{\em Spectral Gap and Exponential Decay of Correlations}\/,
Comm. Math. Phys. {\bf 256}, 781--804 (2006).
\\\url{https://arxiv.org/abs/math-ph/0507008}

\bibitem{NS1}
B. Nachtergaele and R. Sims,
{\em Lieb-Robinson Bounds and the Exponential Clustering Theorem}\/,
Comm. Math. Phys. {\bf 265}, 119--130 (2006).
\\\url{https://arxiv.org/abs/math-ph/0506030}

\bibitem{AKLT1}
I. Affleck, T. Kennedy, E.H. Lieb, and H. Tasaki,
{\em Rigorous results on valence-bond ground states in antiferromagnets}\/,
Phys. Rev. Lett. {\bf 59}, 799 (1987).

\bibitem{AKLT2}
I. Affleck, T. Kennedy, E.H. Lieb, and H. Tasaki,
{\em Valence bond ground states in isotropic quantum antiferromagnets}\/,
Comm. Math. Phys. {\bf 115}, 477--528 (1988).
\\\url{https://projecteuclid.org/euclid.cmp/1104161001}

\bibitem{FannesNachtergaeleWerner1992}
M. Fannes, B. Nachtergaele, and R. F. Werner,
{\em Finitely correlated states on quantum spin chains}\/,
Comm. Math. Phys. {\bf  144}, 443--490 (1992).
\\\url{https://projecteuclid.org/euclid.cmp/1104249404}

\bibitem{denNijsRommelse}
M. den Nijs and K. Rommelse,
{\em Preroughening transitions in crystal surfaces and valence-bond phases in quantum spin chains}\/,
Phys. Rev. B {\bf 40}, 4709 (1989).

\bibitem{KennedyTasaki1992A}
T. Kennedy  and H. Tasaki,
{\em Hidden  $\ZZ$ symmetry breaking in Haldane-gap antiferromagnets}\/,
Phys. Rev. B {\bf 45}, 304--307 (1992).

\bibitem{KennedyTasaki1992B}
T. Kennedy  and H. Tasaki,
{\em Hidden symmetry breaking and the Haldane phase in $S= 1$ quantum spin chains}\/, 
Comm. Math. Phys. {\bf 147}, 431--484 (1992).
\\\url{https://projecteuclid.org/euclid.cmp/1104250747}

\bibitem{Oshikawa92}
M. Oshikawa, 
{\em Hidden $ \mathbb{Z}_2\times \mathbb{Z}_2$ symmetry in quantum spin chains with arbitrary integer spin}\/,
J. Phys. Cond. Matt. {\bf 4}, 7469 (1992).

\bibitem{GuWen2009}
Z.-C. Gu,  and X.-G. Wen,
{\em Tensor-entanglement-filtering renormalization approach and symmetry-protected topological order}\/, 
Phys. Rev. B {\bf 80}, 155131 (2009).
\\\url{https://arxiv.org/abs/0903.1069}

\bibitem{PollmannTurnerBergOshikawa2010}
F. Pollmann, A.M. Turner, E. Berg, and M. Oshikawa,
{\em Entanglement spectrum of a topological phase in one dimension}\/,
Phys. Rev. B {\bf 81}, 064439 (2010).
\\\url{https://arxiv.org/abs/0910.1811}

\bibitem{PollmannTurnerBergOshikawa2012}
F. Pollmann, A.M. Turner, E. Berg, and M. Oshikawa,
{\em Symmetry protection of topological phases in one-dimensional quantum spin systems}\/,
Phys. Rev. B {\bf 85}, 075125 (2012).
\\\url{https://arxiv.org/abs/0909.4059}


\bibitem{ChenGuWEn2011}
X. Chen, Z.-C. Gu, and X.-G. Wen,
{\em Classification of gapped symmetric phases in one-dimensional spin systems}\/,
Phys. Rev. B {\bf 83}, 035107 (2011).
\\\url{https://arxiv.org/abs/1008.3745}

\bibitem{Ogata1}
Y. Ogata,
{\em A class of asymmetric gapped Hamiltonians on quantum spin chains and its characterization I}\/,
Commun. Math. Phys. {\bf 348}, 847--895 (2016).
\\\url{https://arxiv.org/abs/1510.07753}

\bibitem{Ogata2}
Y. Ogata,
{\em A class of asymmetric gapped Hamiltonians on quantum spin chains and its characterization II}\/,
Commun. Math. Phys. {\bf 348}, 897--957 (2016).
\\\url{https://arxiv.org/abs/1510.07751}

\bibitem{Ogata3}
Y. Ogata,
{\em A class of asymmetric gapped Hamiltonians on quantum spin chains and its characterization III}\/,
Commun. Math. Phys. {\bf 352}, 1205--1263 (2017).
\\\url{https://arxiv.org/abs/1606.05508}

\bibitem{OgataZ2}
Y. Ogata,
{\em A $ \mathbb{Z}_2$-index of symmetry protected topological phases with time reversal symmetry for quantum spin chains}\/,
Comm. Math. Phys. {\bf 374}, 705--734 (2020).
\\\url{https://arxiv.org/abs/1810.01045}

\bibitem{Ogatainv}
Y. Ogata,
{\em A $ \mathbb{Z}_2$-index of symmetry protected topological phases with reflection symmetry for quantum spin chains}\/,
to appear in 
Comm. Math. Phys. (2022).
\\\url{https://arxiv.org/pdf/1904.01669.pdf}

\bibitem{OgataCDM}
Y. Ogata,
{\em Classification of symmetry protected topological phases in quantum spin chains}\/,
Proceedings of Current Developments in Mathematics, to appear.
\\\url{https://arxiv.org/abs/2110.04671}

\bibitem{HalSPT}
H. Tasaki,
{\em Symmetry-protected topological (SPT) phases and topological indices in quantum spin chains}\/,
Online lecture (2021).
\\\url{https://www.gakushuin.ac.jp/~881791/OL/index.html#SPT2021}
\\\url{https://youtu.be/xmKA0jwWXec}


\bibitem{BachmannNachtergaele2012}
S. Bachmann and B. Nachtergaele,
{\em Product vacua with boundary states}\/,
Phys. Rev. B {\bf 86}, 035149 (2012).
\\\url{https://arxiv.org/abs/1112.4097v2}

\bibitem{BachmannNachtergaele2014}
S. Bachmann and B. Nachtergaele,
{\em Product vacua with boundary states and the classification of gapped phases}\/,
Commun. Math. Phys. {\bf 329}, 509--544 (2014).
\\\url{https://arxiv.org/abs/1212.3718}

\bibitem{TasakiLSM}
H. Tasaki, {\em The Lieb-Schultz-Mattis Theorem: A Topological Point of View}\/,
in Rupert L. Frank, Ari Laptev, Mathieu Lewin, and Robert Seiringer eds. ``The Physics and Mathematics of Elliott Lieb'' vol.~2, pp.~405--446 (European Mathematical Society Press, 2022).
\\\url{https://arxiv.org/abs/2202.06243}

\bibitem{KLZ}
 A. Kl\"{u}mper, A. Schadschneider, and J. Zittartz,
 {\em Matrix Product Ground States for One-Dimensional Spin-1 Quantum Antiferromagnets}\/,
Europhys. Lett. {\bf 24}, 293--297 (1993).
\\\url{https://arxiv.org/abs/cond-mat/9307028}

\bibitem{KennedyLiebTasaki1988}
T. Kennedy, E.H. Lieb, and H. Tasaki,
{\em A two-dimensional isotropic quantum antiferromagnet with unique disordered ground state}\/,
J. Stat. Phys.  {\bf 53}, 383--415 (1988). 

\bibitem{Knabe}
S. Knabe,
{\em Energy gaps and elementary excitations for certain VBS-quantum antiferromagnets}\/,
J. Stat. Phys. {\bf 52}, 627--638 (1988).

\bibitem{BHM}
S. Bravyi, M.B. Hastings, and S. Michalakis,
{\em Topological quantum order: stability under local perturbations}\/,
J. Math. Phys. {\bf 51}, 093512 (2010).
\\\url{https://arxiv.org/abs/1001.0344}

\bibitem{BH}
S. Bravyi and M.B. Hastings,
{\em A Short Proof of Stability of Topological Order under Local Perturbations}\/,
Commun. Math. Phys. {\bf 307}, 609--627 (2011).
\\\url{https://arxiv.org/abs/1001.4363}

\bibitem{NSY}
B. Nachtergaele, R. Sims, and A. Young,
{\em Quasi-Locality Bounds for Quantum Lattice Systems. Part II. Perturbations of Frustration-Free Spin Models with Gapped Ground States}\/,
Ann. Henri Poincar\'{e} {\bf 23}, 393--511 (2022).
\\\url{https://arxiv.org/abs/2010.15337}




\bibitem{PerezGarciaVerstraete}
D. Perez-Garcia, F. Verstraete, M.M. Wolf, and J.I. Cirac,
{\em Matrix Product State Representations}\/,
Quantum Inf. Comput. 7, 401 (2007).
\\\url{https://arxiv.org/abs/quant-ph/0608197}


\bibitem{Thouless}
D. J. Thouless,
{\em Quantization of particle transport}\/,
Phys. Rev. B {\bf 27}, 6083 (1983).

\bibitem{Shindou}
R. Shindou,
{\em Quantum Spin Pump in $S=1/2$ Antiferromagnetic Chains: Holonomy of Phase Operators in sine-Gordon Theory}\/,
J. Phys. Soc. Jpn. {\bf 74}, 1214--1223 (2005).
\\\url{https://journals.jps.jp/doi/10.1143/JPSJ.74.1214}

\bibitem{KunoHatsugai2021}
Y. Kuno and Y. Hatsugai,
{\em Plateau transitions of a spin pump and bulk-edge correspondence}\/,
Phys. Rev. B {\bf 104}, 045113 (2021).
\\\url{https://arxiv.org/abs/2102.09325}


\bibitem{Kapustin2020}
A. Kapustin and L. Spodyneiko,
{\em Higher-dimensional generalizations of Berry curvature}\/,
Phys. Rev. B {\bf 101}, 235130 (2020).
\\\url{https://arxiv.org/abs/2001.03454}

\bibitem{Wen2021}
X. Wen, M. Qi, A. Beaudry, J. Moreno, M.J. Pflaum, D. Spiegel, A. Vishwanath, and M. Hermele,
{\em Flow of (higher) Berry curvature and bulk-boundary correspondence in parametrized quantum systems}\/,
preprint (2021).
\\\url{https://arxiv.org/abs/2112.07748}

\bibitem{Bachmann2022}
S. Bachmann, W. De Roeck, M. Fraas, and T. Jappens,
{\em A classification of G-charge Thouless pumps in 1D invertible states}\/,
preprint (2022).
\\\url{https://arxiv.org/abs/2204.03763}

\bibitem{TasakiLiebConference}
H. Tasaki,
{\em Variations on a Theme by Lieb, Schultz, and Mattis: Unique Gapped Ground States, Symmetry-Protected Topological Phases, Edge States, Spin Pumping, and all that in Quantum Spin Chains}\/,
Online lecture (2022).
\\\url{https://www.gakushuin.ac.jp/~881791/OL/index.html#varLSM}
\\\url{https://youtu.be/XUBicfQN6kk}

\bibitem{TasakiNext}
H. Tasaki, {\em Topological Indices, Symmetry Protected Topological Phases, Gapless Edge Excitations, Spin Pumping, and Homotopy in Quantum Spin Chains}\/, in preparation.

\bibitem{Bohm}
D. Bohm, {\em Note on a theorem of Bloch concerning possible causes of superconductivity}\/, Phys. Rev. {\bf 75},
502 (1949).

\bibitem{LSM}
E. Lieb, T. Schultz, and D. Mattis,
{\em Two soluble models of an antiferromagnetic chain}\/,
Ann. Phys. {\bf 16}, 407--466 (1961).

\bibitem{Witten}
E. Witten, 
{\em Constraints on supersymmetry breaking}\/, 
Nucl. Phys. B 202, 253 (1982).

\bibitem{WoutersKatsuraSchuricht}
J. Wouters, H. Katsura, and D. Schuricht,
{\em Interrelations among frustration-free models via Witten's conjugation}\/,
SciPost Phys. Core 4, 027 (2021).
\\\url{https://arxiv.org/abs/2005.12825}


\bibitem{ASZ}
M.A. Ahrens, A. Schadschneider, and J. Zittartz,
{\em Exact ground states of spin-2 chains}\/,
Europhys. Lett. {\bf 59}, 889--895 (2002).
\\\url{https://arxiv.org/abs/cond-mat/0206537}




\end{thebibliography}
\end{document}